\catcode`\@=11
\def\citen#1{\if@filesw \immediate\write \@auxout {\string\citation{#1}}\fi%
\@tempcntb\m@ne \let\@h@ld\relax \def\@citea{}%
\@for \@citeb:=#1\do {\@ifundefined {b@\@citeb}%
    {\@h@ld\@citea\@tempcntb\m@ne{\bf ?}%
    \@warning {Citation `\@citeb ' on page \thepage \space undefined}}%
    {\@tempcnta\@tempcntb \advance\@tempcnta\@ne
    \setbox\z@\hbox\bgroup\ifcat0\csname b@\@citeb \endcsname \relax
    \egroup \@tempcntb\number\csname b@\@citeb \endcsname \relax
    \else \egroup \@tempcntb\m@ne \fi \ifnum\@tempcnta=\@tempcntb
    \ifx\@h@ld\relax \edef \@h@ld{\@citea\csname b@\@citeb\endcsname}%
    \else \edef\@h@ld{\hbox{--}\penalty\@highpenalty
    \csname b@\@citeb\endcsname}\fi
    \else \@h@ld\@citea\csname b@\@citeb \endcsname \let\@h@ld\relax \fi}%
\def\@citea{,\penalty\@highpenalty\hskip.13em plus.13em minus.13em}}\@h@ld}
\def\@citex[#1]#2{\@cite{\citen{#2}}{#1}}%
\def\@cite#1#2{\leavevmode\unskip\ifnum\lastpenalty=\z@\penalty\@highpenalty\fi%
  \ [{\multiply\@highpenalty 3 #1%
  \if@tempswa,\penalty\@highpenalty\ #2\fi}]}   %
\makeatother 

\catcode`\@=12

\def\Ad            {{\rm Ad}}
\def\aff           {affine Lie algebra}
\def\alg           {algebra}
\def\auto          {automorphism}

\newcommand\Barray[2]{\mbox{\large$[$}\!\!{\scs\begin{array}{c}{}\\[-1.94em]
                   {\scs #1}\\[-.43em]{\scs #2}\\[-.4em] \end{array}}\!\!%
                   \mbox{\large$]$}}
\def\bc            {boundary condition}
\def\Bc            {Boundary condition}
\def\be            {\begin{equation}}
\def\bearl         {\begin{array}{l}}
\def\bearll        {\begin{array}{ll}}
\def\bearlll       {\begin{array}{lll}}
\def\bfe           {{\bf1}}
\def\bP            {\begin{picture}}
\def\bPo           {\begin{picture}(0,0)}
\def\cala          {{\mathfrak A}}
\def\calap         {{\bar\cala}}
\def\calb          {{\cal B}}
\def\calc          {{\cal C}}
\def\calco         {\calc_\circ}
\def\calf          {{\cal F}}
\def\calh          {{\cal H}}
\def\call          {{\cal L}}
\def\caln          {{\cal N}}
\def\calhb         {\bar\calh}
\def\calz          {{\cal Z}}
\def\cft           {conformal field theory}
\def\Cft           {Conformal field theory}
\def\cfts          {conformal field theories}

\def\chii          {\raisebox{.15em}{$\chi$}}
\def\complex       {{\dl C}}

\def\csa           {Cartan subalgebra}
\def\dl            {\mathbb }
\def\dotomego      {{\dot\omega_\circ}}
\def\DP            {{D\Pa}}
\def\dsty          {\displaystyle}
\def\dyd           {Dynkin diagram}
\def\ee            {\end{equation}}
\def\eE            {{\rm e}}
\def\eear          {\end{array}}
\def\eP              {\end{picture}}
\def\EP            {{E\Pa}}
\def\eps           {\epsilon}
\def\eq            {\,{=}\,}
\newcommand\erf[1] {(\ref{#1})}
\newcommand\Erf[2] {(\ref{#1#2})}
\newcommand\Erra[3]{\,[{\em ibid.}\ {#1} ({#2}) {#3}, {\em Erratum}]}
\newcommand\F[6]   {{\sf F}_{\! #1,#2}\Barray{#4\,#5}{#3\,#6}}
\newcommand\fc[4]{{#1}\hspace{-#2em}\raisebox{#3em}{$\scriptstyle\circ$}
                   \hspace{#4em}}
\def\FF            {{\sf F}}
\def\findim        {finite-dimensional}
\newcommand\Frac[2]{\mbox{\large$\frac{#1}{#2}$}}
\def\furu          {fusion rule}
\def\futnote#1     {\footnote{~#1}\ }
\def\g             {{\mathfrak g}}
\def\G             {{\rm G}}
\def\gb            {\bar\g}

\def\gv            {g^\Vee}
\newcommand\hsp[1] {\mbox{\hspace{#1 em}}}
\def\hy            {$\mbox{-\hspace{-.66 mm}-}$}
\def\id            {\mbox{\sl id}}
\def\ii            {{\rm i}}
\def\iN            {\,{\in}\,}

\def\Infdim        {Infinite-dimensional }
\newcommand\infobox[2] {\framebox(27,21){$\begin{array}c{}\\[-2.2em]\sss
                   x\,=\,#1\\[-.42em]\sss k\,=\,#2 {}\\[-.97em] \eear$}}

\def\irmod         {irreducible module}

\def\kma           {Kac\hy Moody algebra}
\def\KP            {{K\Pa}}
\long\def\labl#1   {\label{#1}\ee}
\long\def\Labl#1#2 {\label{#1#2}\ee}
\def\lie           {Lie algebra}
\def\Lie           {Lie group}
\def\llb           {\mbox{\large(}}
\def\LP            {{L\Pa}}
\def\lrb           {\mbox{\large)}}
\def\Lv            {L^{\!\vee}_{\phantom|}}
\def\LV            {L\raisebox{.51em}{$\sss\!\vee$}}
\def\LVomega       {L\raisebox{.51em}{$\sss\!\vee$}{}_{\!\!\omega}}
\def\LVomegas      {M\raisebox{.51em}{$\sss{\rm w}$}{}_{\!\!\!\!\omega}}
\def\Lw            {L\raisebox{.51em}{${\sss\!{\rm w}}$}}
\def\Lwomega       {L\raisebox{.51em}{${\sss\!{\rm w}}$}{}_{\!\!\omega}}
\def\Lwomegas      {M\raisebox{.51em}{$\sss\vee$}{}_{\!\!\!\!\omega}}
\def\m             {{\rm mult}}
\def\mimo          {minimal model}
\def\modinv        {modular invarian}
\newcommand\N[3]   {{\rm N}_{#1 #2}^{\ \ #3}}
\def\oa            {operator algebra}
\def\omegas        {{\omega^\star}}
\def\omego         {{\omega_\circ}}
\def\one           {\mbox{\small $1\!\!$}1}
\def\onedim        {one-dimen\-sional}
\def\ot            {\raisebox{.07em}{$\scriptstyle\otimes$}}
\def\oT            {\,\ot\,}
\def\otimeS        {\,{\otimes}\,}
\def\P             {^+_{\phantom i}}
\def\Pa            {}
\def\parfu         {partition function}
\def\PP            {^{\!+}_{\phantom i}}
\newcommand\pso[4] {\Psi_{\;#1}^{#2\,{\sss #4}\,#3}}
\newcommand\pSchannel[9]{\bP(115,80)(#1,#2)
                   \Schannel{#3}{#4}{#5}{#6}{#7}{#8}{#9}\eP}
\newcommand\Schannel[7]{ \putlin{50}{50}10{30} \putlin{50}{50}{-1}1{22}
                   \putlin{50}{50}{-1}{-1}{22} \putlin{80}{50}11{22}
                   \putlin{80}{50}1{-1}{22} \putvec{70}{50}10
                   \putvec{30}{70}1{-1} \putvec{30}{30}11
                   \putvec{100}{70}{-1}{-1} \putvec{100}{30}{-1}1
                   \putsc{19.6}{20.5}{#1} \putsc{19.6}{75}{#2}
                   \putsc{106}{75}{#3} \putsc{106}{20.5}{#4}
                   \putsc{63}{51.5}{#5} \putss{39.7}{48.4}{#6}
                   \putss{83.9}{48.4}{#7} }
\newcommand\pTchannel[9]{\bP(77,80)(#1,#2)
                   \Tchannel{#3}{#4}{#5}{#6}{#7}{#8}{#9}\eP}
\newcommand\Tchannel[7]{ \putlin{30}{50}01{30} \putlin{30}{80}11{22}
                   \putlin{30}{80}{-1}1{22} \putlin{30}{50}1{-1}{23}
                   \putlin{30}{50}{-1}{-1}{23} \putvec{30}{60}0{-1}
                   \putvec{10}{100}1{-1} \putvec{10}{30}11
                   \putvec{50}{100}{-1}{-1} \putvec{50}{30}{-1}1
                   \putsc{0.5}{19}{#1} \putsc{0.5}{104}{#2}
                   \putsc{54.5}{104}{#3} \putsc{54.5}{19}{#4}
                   \putsc{32.5}{63}{#5} \putss{27.7}{84}{#6}
                   \putss{27.7}{42.5}{#7} }
\newcommand\putlin[5]{\put(#1,#2){\line(#3,#4){#5}}}

\newcommand\putsc[3] {\put(#1,#2){{$\scriptstyle #3$}}}

\newcommand\putss[3] {\put(#1,#2){{$\scriptscriptstyle #3$}}}
\newcommand\putvec[4]{\put(#1,#2){\vector(#3,#4)5}}
\def\q             {quantum }
\def\Q             {Quantum }
\def\rep           {representation}

\def\rmd           {{\rm d}}
\def\scs           {\scriptstyle}
\newcommand\sect[1]{\section{#1}\setcounter{equation}{0}}
\def\SO            {{\rm SO}}
\def\sss           {\scriptscriptstyle}
\def\SU            {{\rm SU}}
\newcommand\suma[4]{\sum_{#1=1}^{\N{#3}{#4}{#2}}}
\newcommand\sumn[4]{\sum_{#1=1}^{\N{#2}{#3}{#4}}}
\newcommand\sxs[3] {{{{\sss #1}}{#2}{{\sss #3}}}}
\def\syms          {sym\-me\-tries}
\newcommand\tC[9]  {C_{\,#1\,{\sss #6}\,#2}
                   ^{#3\,{\sss #7}\,#4\,{\sss #8}\,#5\,{\sss #9}}}
\newcommand\tCv[7] {C_{\,#1\,#2\;\vac}^{#3\,{\sss #6}\,#4\,{\sss #7}\,#5}}
\def\tr            {{\rm tr}}
\def\twodim        {two-dimensional}
\def\vac           {\Omega}
\def\Vee           {{\sss\vee}}

\def\vphi          {\varphi}
\def\wp            {\fc w{.5}{.48}{.1}}
\def\wP            {\fc w{.37}{.32}{.1}}
\def\Wp            {\fc W{.7}{.78}{.2}}
\def\WP            {\fc W{.55}{.52}{.2}}

\def\wrtt          {with respect to the }
\def\WZW           {Wess\hy Zu\-mino\hy Wit\-ten }
\def\wzwm          {WZW model}
\def\wzwt          {WZW theory}
\def\wzwts         {WZW theories}
\def\zet           {{\dl Z}}
\def\zetplus       {{\dl Z}_{>0}}

\documentclass[12pt]{article} \usepackage{amssymb,amsfonts,latexsym}
\usepackage{epsfig}

\begin{document}

\begin{flushright}  {~} \\[-1cm] {\sf hep-th/9909030} \\[1mm]
{\sf ETH-TH/99-24} \\[1 mm] {\sf ESI-751} \\[1 mm]
{\sf September 1999} \end{flushright}
 
\begin{center} \vskip 22mm
{\Large\bf THE GEOMETRY OF WZW BRANES}\\[22mm]
{\large Giovanni Felder}\,, \ {\large J\"urg Fr\"ohlich}\,, \ 
{\large J\"urgen Fuchs} \ and \ {\large Christoph Schweigert}\\[5mm]
ETH Z\"urich\\[.2em] CH -- 8093~~Z\"urich
\end{center}
\vskip 26mm
\begin{quote}{\bf Abstract}\\[1mm]
The structures in target space geometry that correspond to conformally 
invariant boundary conditions in WZW theories are determined both by 
studying the scattering of closed string states and by investigating 
the algebra of open string vertex operators. In the limit of large level, 
we find branes whose world volume is a regular conjugacy class or, in the 
case of symmetry breaking boundary conditions, a `twined' version thereof.
In particular, in this limit one recovers the commutative algebra of functions 
over the brane world volume, and open strings connecting different branes
disappear. At finite level, the branes get smeared out, yet their approximate 
localization at (twined) conjugacy classes can be detected unambiguously.
\\
As a by-product, it is demonstrated how the pentagon identity and 
tetrahedral symmetry imply that in any rational conformal field theory the 
structure constants of the algebra of boundary operators coincide with 
specific entries of fusing matrices. 
\end{quote}
\newpage


\sect{Introduction}

Conformally invariant boundary conditions in \twodim\ \cfts\ have recently 
attracted renewed attention. By now, quite a lot of information on such 
boundary conditions is available in the algebraic approach, including
boundary conditions that do not preserve all bulk symmetries.
In many cases, the \cft\ of interest has also a description as a 
sigma model with target space $M$. It is then tempting to ask
what the geometrical interpretation of these boundary conditions might
be in terms of submanifolds (and vector bundles on them or, more generally, 
sheaves) of $M$. Actually, this question makes an implicit assumption 
that is not really justified: It is not the classical (commutative) 
geometry of the target $M$ that matters, but rather 
a non-commutative version \cite{frgA} of it. 

In the present note, we investigate the special case of WZW conformal
field theories.  For most of the time we restrict our attention to the 
case where the torus \parfu\ is given by charge conjugation. Then the 
classical target space is a real simple compact connected and simply 
connected Lie group manifold $G$. In particular, the underlying manifold
is parallelizable, i.e.\ its tangent bundle is a trivial bundle. 

The latter property of Lie groups will allow us to apply methods that were
developped in \cite{dfpslr}, by which geometric features of the D-brane
solutions of supergravity in flat 10-dimensional space-time were recovered
from the boundary state for a free conformal field theory. The basic idea of 
that approach was to compute the vacuum expectation value of the {\em bulk\/} 
field that corresponds to the {\em closed\/} string state
  \be  \alpha^\mu_{-1}\, \tilde\alpha^\nu_{-1} |q\rangle \ot |\tilde q\rangle
  \labl4
on a disk with a boundary condition $\beta$ of interest. Here our convention
is that quantities without a tilde correspond to left-movers, while those with
a tilde correspond to right-movers. The operator $\alpha^\mu_n$ is the $n$th 
mode of the $\mathfrak u(1)$ current in the $\mu$-direction of the free \cft. 
The symmetric traceless part of the state \erf4 corresponds to the graviton, 
the antisymmetric part of the state to the Kalb\hy Ramond field, and the 
trace to the dilaton, all of momentum $q$.

Let us explain the rationale behind this prescription. At first sight it might 
seem more natural to employ graviton scattering in the background of a brane 
for exploring the geometry. This would correspond, in leading order of string
perturbation theory, to the calculation of a two-point correlation function 
for two bulk fields on the disk. However, by factorization of bulk fields 
such an amplitude is related to (a sum of) products of three-point functions
on the sphere with one-point functions on the disk. Since the former amplitude 
is completely independent
of the boundary conditions, all information on a boundary condition that
can be obtained by use of {\em bulk\/} fields will therefore be obtainable from
correlators involving a single bulk field. Similar factorization arguments also
encourage us to concentrate on world sheets with the topology of a disk.

The idea of testing boundary conditions with vacuum expectation values of
bulk fields finds an additional justification in the following reasoning. 
In terms of classical geometry, boundary conditions are related to vector 
bundles over submanifolds of the target manifold $M$, the Chan\hy Paton bundles.
Such bundles, in turn, should be regarded as modules over the ring $\calf(M)$
of functions on $M$. Heuristically, we may interpret the algebra of (certain)
bulk fields as a quantized version of $\calf(M)$. The expectation values of 
the bulk fields on a disk then describe how the algebra of bulk fields is
represented on the boundary operators or, more precisely, on the subspace of
boundary operators that are descendants of the vacuum field. (As a side remark 
we mention that boundary conditions are indeed most conveniently described in 
terms of suitable {\em classifying algebras\/}. These encode aspects of the 
action of the algebra of bulk operators on boundary operators.)

The relevant information for computing the one-point functions on a disk with
boundary condition $\beta$ is encoded in a {\em boundary state\/} 
$\calb_\beta$, which is a linear functional 
  \be  \calb_\beta: \quad \bigoplus_{q,\tilde q} \, 
  \calh_q\otimeS \calh_{\tilde q} \to \complex \ee
on the space of closed string states. (For an uncompactified free boson, the 
left- and right-moving labels of the primary fields are related as 
$\tilde q\eq q$.) We are thus led to compute the function
  \be  \G_\beta^{\mu\nu}(q) := \calb_\beta(\alpha^\mu_{-1} \tilde\alpha^\nu_{-1} 
  |q\rangle \ot |\tilde q\rangle) \,. \labl2
Using the explicit form of the boundary state, this quantity has been 
determined in \cite{dfpslr}. Upon Fourier transformation, it gives rise to
a function $\tilde\G_\beta^{\mu\nu}(x)$ on position space. It has been
shown that the symmetric traceless part of the function $\G_\beta$ 
reproduces the vacuum expectation value of the graviton in the background of 
a brane, while the antisymmetric part gives the Kalb\hy Ramond field, and the 
trace the dilaton.

In order to see how these findings generalize to the case of (non-abelian)
\wzwts, let us examine the structural ingredients that enter in these
calculations. Boundary states can be constructed for arbitrary \cfts, in 
particular for \wzwm s. Moreover, since group manifolds are parallelizable, it 
is also straightforward to generalize the oscillator modes $\alpha^\nu_n$: 
they are to be replaced by the corresponding modes $J^a_n$ of the non-abelian 
currents $J^a(z)$. Here the upper index $a$ ranges over a basis of the \lie\ of 
$G$, $a\eq1,2,...\,,\dim G$, and $n\iN\zet$. Together with a central element 
$K$, these modes span an untwisted affine \lie\ $\g$, according to
  \be [J^a_n, J^b_m] = \sum_c f^{ab}_{\;\ c}\, J^c_{n+m}
  + n \,\kappa^{ab} \delta_{n+m,0}\, K  \,.  \labl{aff}
Here $f^{ab}_{\;\ c}$ and $\kappa^{ab}$ are the structure constants and
Killing form, respectively, of the \findim\ simple \lie\ $\gb$ whose
compact real form is the Lie algebra of the Lie group manifold $G$.
Notice that the generators of the form $J_0^a$ form a \findim\ subalgebra,
called the horizontal subalgebra, which can (and will) be identified with $\gb$.

We finally need to find the correct generalization of the state $|q\rangle$. 
To this end we note that $|q\rangle$ is the vector in the Fock space of charge 
$q$ with lowest conformal weight. For \wzwts, instead of this Fock space, we 
have to consider the following space. First, we must choose a non-negative 
integer value $k$ for the level, i.e., the eigenvalue of the central element $K$.
The space of physical states of the \wzwt\ with charge conjugation modular
invariant is then the direct sum
  \be \bigoplus_{\lambda\in P_k} \calh_\lambda^{}\otimeS \calh_{\lambda\PP}
  \,,  \ee
where $\calh_\lambda$ is the irreducible integrable highest weight 
module of $\g$ at level $k$ with highest weight $\lambda$, and 
$P_k$ is the (finite) set of integrable weights $\lambda$ of $\g$ at 
level $k$. Every such $\g$-weight $\lambda$ corresponds to a unique 
weight of the horizontal subalgebra $\gb$ (which we denote again by
$\lambda$), which is the highest weight of a \findim\ $\gb$-\rep.
However, for finite level $k$, not all such highest weights of 
$\gb$ appear; this truncation will have important consequences later on.

Unlike in the case of Fock modules, the subspace of states of lowest conformal
weight in the module $\calh_\lambda$ is not \onedim\ any longer. Rather,
it constitutes the irreducible \findim\ module
$\calhb_\lambda$ of the horizontal subalgebra $\gb$.
Therefore in place of the function \erf2 we now consider
  \be  \G_\beta^{ab}(v\ot\tilde v) := 
  \calb_\beta(J^a_{-1} \tilde J^b_{-1}\, |v\rangle \ot |\tilde v\rangle)
  \Labl,G
for 
  \be v\oT \tilde v\in \bigoplus_{\lambda\in P_k}
  \calhb_{\lambda}^{}\otimeS \calhb_{\lambda\PP} \,.  \labl3
As a matter of fact, one may also look at analogous quantities involving other 
modes $J^a_n$, or combinations of modes, or even without any mode present
at all. It turns out that qualitatively their behavior is very similar
to the functions \Erf,G; they all signal the presence of a defect at the 
same position in target space. Our results are therefore largely independent
on the choice of the bulk field we use to test the geometry of the target.

The function $\G_\beta^{ab}$ can be determined from known results about \bc s 
in \wzwm s. This allows us to analyze WZW brane geometries via expectation
values of bulk fields. Another approach to these geometries is via the \alg\ 
of boundary fields. While the second setup focuses on intrinsic properties
of the brane world volume, the first perspective offers a natural way to
study the embedding of the brane geometry into 
the target. Both approaches will be studied in this paper. 

We organize our discussion as follows. In section 2 we compute the function 
$\G_\beta^{ab}$ for those boundary conditions which preserve all bulk 
symmetries. To relate this function to classical geometry of the group manifold 
$G$, we perform a Fourier transformation. We then find that the end points of
open strings are naturally localized at certain conjugacy classes of
the group $G$. At finite level $k$, the locus of the end points of the open
string is, however, smeared out, though it is still well peaked at a definite
regular rational conjugacy class. The absence of sharp localization at finite 
level $k$ shows that, even after having made the relation to classical geometry, 
the brane exhibits some intrinsic `fuzziness'. It should, however, be emphasized 
that at finite level the very concept of both the target space and the world 
volume of a brane as classical \findim\ manifolds are not really appropriate. 

The algebra of boundary fields for symmetry preserving boundary conditions
is analyzed in section 3. It can be shown that for any arbitrary rational \cft\ 
the boundary structure constants are equal to world sheet duality matrices, the 
fusing matrices, according to
  \be  \tC{\lambda\mu}\nu\alpha\beta\gamma LABC
  = \llb \F{\sxs L\nu C}{\sxs A{\beta\P}B}{\alpha\P}{\lambda\;}{\;\mu_{}}
  \gamma {\lrb}^* \,. \Labl CF
Furthermore, we are able to show that in the limit of large $k$ the 
space of boundary operators approaches the space of functions on the
brane world volume. In the same limit open strings connecting different
conjugacy classes disappear, while such configurations are present
at every finite value of the level.

In section 4 we discuss symmetry breaking boundary conditions in \wzwts\ for
which the symmetry breaking is characterized through an automorphism $\omega$
of the horizontal subalgebra $\gb$.\,%
 \futnote{Not all symmetry breaking boundary conditions of
\wzwts\ are of this form.}
 It turns out that the end points of open strings are then localized
at {\em twined conjugacy classes\/}, that is, at sets of the form
  \be   \calc_G^\omega(h) := \{ g h \omega(g)^{-1} \,|\, g\iN G \}  \ee
for some $h\iN G$. The derivation of our results on symmetry breaking boundary 
conditions requires generalizations of Weyl's classical results on conjugacy 
classes. (The necessary tools, including a twined version of Weyl's 
integration formula, are collected in appendix B.) In section 5, we extend 
our analysis to non-simply connected Lie groups. We find features that are 
familiar from the discussion of D-branes on orbifold spaces, such as fractional
branes, and point out additional subtleties in cases where the action of 
the orbifold group is only projective.

\sect{Probing target geometry with bulk fields}

We start our discussion with the example of boundary conditions that preserve
all bulk symmetries. In this situation the correlators on a surface with
boundaries are specific linear combinations of the chiral blocks on
the Schottky double of the surface \cite{fuSc6}. The boundary state describes 
the one-point correlators for bulk fields on the disk and, accordingly, it
is a linear combination of two-point blocks on the Schottky cover of the 
disk, i.e., on the sphere. The latter -- which in the present context of
correlators on the disk also go under the name of Ishibashi states --
are linear functionals
  \be  B_\lambda \,: \quad \calh_\lambda^{}\otimeS\calh_{\lambda\PP}\to\complex
  \ee
that are characterized by the Ward identities
  \be  B_\lambda \circ \left( J^a_{n}\oT\bfe + \bfe\oT J^a_{-n} \right)
  = 0 \,.  \labl1
Choosing an element $v\ot\tilde v\iN \calhb_\lambda^{}{\otimes}
\calhb_{\lambda\PP}$, we can use the invariance property \erf1 and the 
commutation relations \erf{aff} to arrive at
  \be \bearll
  B_\lambda(J^a_{-1} v \oT J^b_{-1} \tilde v) \!\!
  &= - B_\lambda((\bfe\oT J^a_{1} J^b_{-1})\,(v\ot \tilde v)) 
   = - B_\lambda((\bfe\oT [J^a_1,J^b_{-1}])\,(v\ot \tilde v)) 
  \\{}\\[-.6em]
  &= - \dsty\sum_c f^{ab}_{\;\ c}\, B_\lambda(v\oT J^c_0\tilde v)
  - \kappa^{ab} k\, B_\lambda(v\ot \tilde v) \,.
  \end{array}\ee

There is one symmetry preserving boundary condition for each primary
field $\alpha$ in the theory \cite{card9}. The coefficients in the 
expansion of the boundary states \wrtt boundary blocks
are given by the so-called (generalized) quantum dimensions:
  \be  \calb_\alpha = \sum_{\lambda\in P_k} \frac{S_{\lambda,\alpha}}
  {S_{\Omega,\alpha}}\, B_\lambda \,.  \Labl,1
Here $S$ is the modular S-matrix of the theory and $\Omega$ refers to 
the vacuum primary field. To write the state \Erf,1 in a more convenient
form, we use the fact that the generalized quantum dimensions are given by 
the values of the $\gb$-character of $\calhb_\lambda$
at specific elements $y_\alpha$ of the Cartan subalgebra
of the horizontal subalgebra $\gb$ or, equivalently, by the values of the 
$G$-character of $\calhb_\lambda$ at specific elements $h_\alpha$ of 
the maximal torus of the group $G$. Concretely, we have
  \be  S_{\lambda,\alpha} / S_{\Omega,\alpha} = \chii_\lambda(h_\alpha)
  \Labl25
with
  \be  \chii_\lambda(h) := \tr_{\calhb_\lambda} R_\lambda(h)  \ee
and
  \be  h_\alpha := \exp(2\pi\ii y_\alpha) \quad{\rm with}\quad
  y_\alpha := \Frac{\alpha+\rho}{k+\gv} \labl{rcw}
for any level $k$ and $\gb$-weight $\alpha$. In formula \erf{rcw},
$\rho$ denotes the Weyl vector (i.e., half the sum of all positive roots)
of $\gb$ and $\gv$ is the dual Coxeter number. The boundary state thus reads
  \be  \calb_\alpha = \sum_{\lambda\in P_k} \chii_\lambda(h_\alpha)\,
  B_\lambda \,. \ee
The function $\G_\alpha^{ab}$ defined in formula \Erf,G is then found to be 
  \be  \G^{ab}_\alpha(v\ot\tilde v) = - \chii_\lambda(h_\alpha)
  \left[ B_\lambda(v\oT[J^a_0,J^b_0]\tilde v) 
  + \kappa^{ab} k\, B_\lambda(v\ot\tilde v) \right] \ee
for $v\ot\tilde v\iN\calhb_\lambda^{}{\otimes}\calhb_{\lambda\PP}$.

We recall that the group character $\chii$ is a class function, i.e.\ a 
function that is constant on the conjugacy classes
  \be  \calc_G(h) := \{ ghg^{-1} \,|\, g\iN G \}  \Labl cg
of $G$. It is therefore quite natural to associate to a symmetry preserving 
boundary condition the conjugacy class $\calc_G(h_\alpha)$ of the Lie group 
$G$ that contains the element $h_\alpha$. We would like to emphasize that 
$\calc_G(h_\alpha)$ is always a
{\em regular\/} conjugacy class, that is, the stabilizer of $h_\alpha$
under conjugation is just the unique maximal torus containing this element.

Our next task is to perform the analogue of the Fourier transformation
between $\G^{\mu\nu}$ and $\tilde\G^{\mu\nu}$ in \cite{dfpslr}. 
To this end we employ the fact that left and right translation on the group
manifold $G$ give two commuting actions of $G$ on the space $\calf(G)$ of 
functions on $G$ and thereby turn this space into a $G$-bimodule. By the
Peter\hy Weyl theorem, $\calf(G)$ is isomorphic, as a $G$-bimodule, to 
an infinite direct sum of tensor products of \irmod s, namely
  \be  \calf(G) \,\cong\, \bigoplus_{\lambda\in P} \calhb_\lambda^{}\otimes
  \calhb_{\lambda\PP} \,.  \Labl1i
Here $P\,{\equiv}\,P_{k=\infty}$ is the set of all highest weights of \findim\
irreducible $\gb$-modules. We may identify the conjugate module 
$\calhb_{\lambda\PP}$ with the dual of $\calhb_\lambda$. Then the isomorphism 
\Erf1i sends $v\ot\tilde v\iN\calhb_\lambda^{}{\otimes}\calhb_{\lambda\PP}$
to the function $f$ on $G$ given by
  \be  f(g) = \tilde v(R_\lambda(g) v) \equiv
  \langle \tilde v| R_\lambda(g)|v\rangle \ee 
for all $g\iN G$. Using the scalar product on $\calf(G)$, we can therefore
associate to every linear functional $\beta{:}\;\calf\,{\to}\,\complex$ a 
function (respectively, in general, a distribution) $\tilde\beta$
on the group manifold by the requirement that 
  \be  \beta(v\ot\tilde v) = \int_G\!\rmd g\, \tilde\beta(g)^*_{}\,
  \langle \tilde v| R_\lambda(g)|v\rangle\ee
for $v\ot\tilde v\iN\calhb_\lambda^{}{\otimes}\calhb_{\lambda\PP}$.
After introducing dual bases $\{v_i\}$ 
of $\calhb_\lambda^{}$ and $\{\tilde v_j\}$ of $\calhb_{\lambda\PP}$, the 
orthonormality relations for representation functions then allow us to write
  \be  \tilde\beta(g) = \sum_{\lambda\in P} \sum_{i,j} \beta(v_i\ot\tilde
  v_j)^*\, \langle \tilde v_j | R_\lambda(g) | v_i\rangle \, . \ee

According to \erf3, at finite level we have to deal with the
\findim\ truncations 
  \be  \calf_k(G) := \bigoplus_{\lambda\in P_k} 
  \calhb_\lambda^{}\otimeS\calhb_{\lambda\PP}  \Labl1t
of the space \Erf1i of functions on $G$. For every $k$, the space $\calf_k(G)$ 
can be regarded as a subspace of $\calf(G)$. We will do so from now on;
thereby we arrive at a picture that is close to classical intuition. The 
level-dependent truncation \Erf1t constitutes, in fact, one of the basic 
features of a WZW \cft. (This is a typical effect in interacting rational 
\cfts, which does not have an analogue for flat backgrounds.)

Next we relate the linear function $\G^{ab}$ on $\calf_k$ to a 
function $\tilde\G^{ab}$ on the group manifold $G$ by the prescription
  \be  \tilde\G^{ab}(g) := \sum_{\lambda\in P_k} \sum_{i,j}\G^{ab}
  (v_i\ot \tilde v_j)^*_{}\, \langle \tilde v_j | R_\lambda(g) | v_i\rangle \ee
for $g\iN G$. By direct calculation we find
  \be \bearll
  \G^{ab}(g) \!\! &= - \dsty\sum_{\lambda\in P_k} \llb \kappa^{ab} k\,
  \Frac{S_{\lambda,\alpha}^*}{S_{\Omega,\alpha}}\,\tr_{\calhb_\lambda}R_\lambda(g)
  + \sum_c f^{ab}_{\;\ c}\, \Frac{S_{\lambda,\alpha}^*}{S_{\Omega,\alpha}}\,
  \tr_{\calhb_\lambda} J^c R_\lambda(g) \lrb
  \\{}\\[-.8em] 
  &= \dsty -\kappa^{ab} k \sum_{\lambda\in P_k} \chii_\lambda(h_\alpha)^*\,
  \chii_\lambda(g) - \sum_{\lambda\in P_k} \chii_\lambda(h_\alpha)^*\,
  \tr_{\calhb_\lambda} [J^a,J^b] R_\lambda(g)
  \,. \end{array} \Labl2a
In analogy with the situation for flat backgrounds \cite{dfpslr} we are
led to the following interpretation of this result. The first term in the
expression \Erf2a is symmetric and hence describes the vacuum expectation 
value of dilaton and metric that is induced by the presence of the brane, 
while the second term, which is antisymmetric, corresponds to
the vacuum expectation value for the Kalb\hy Ramond field.

To proceed, we introduce, for every $k\iN\zetplus$, a function $\vphi_k$ on 
$G\,{\times}\,G$ by
  \be  \vphi_k(g,h) := \sum_{\lambda\in P_k} \chii_\lambda(g)^{}\,
  \chii_\lambda(h)^* \,.  \ee
In the limit $k\,{\to}\,\infty$ the integral operator associated 
to $\vphi_k$ reduces to the $\delta$-distribution on the space of 
conjugacy classes. Indeed, because of $\lim_{k\to\infty}P_k\eq P$, 
for every class function $f$ on $G$ we have
  \be  \int_G\! \rmd h\, \lim_{k\to\infty} \vphi_k(g,h)\, f(h) = f(g) \,,
  \ee
which is a consequence of the general relation
  \be  \int_G\! \rmd g\, \chii_V(g)^*_{}\, \chii_W(g)
  = \dim({\rm Hom}_G(V,W))  \ee
valid for any two $G$-modules $V,\,W$.
Comparison with the result \Erf2a thus shows that, in the limit of 
infinite level, the brane is localized at the conjugacy class 
$\calc_G(h_\alpha)$ of $G$. It is worth emphasizing, however, that 
this holds true only in that limit. In contrast, at finite level, the 
brane world volume is {\em not\/} sharply localized on the relevant 
conjugacy class $\calc_G(h_\alpha)$. Rather, it gets smeared out or, in more
fancy terms, its localization is on a `fuzzy' version of a conjugacy class.
Nevertheless, already at very small level the localization is 
sufficiently sharp to indicate unambiguously what remains in the limit.

For concreteness, we display a few examples for \bc s with 
$\g\eq\mathfrak{sl}(2)$ in figure 1. The functions of interest are
  \be  f_{k,\alpha}(h) := \caln\,J(h)\,\mbox{\Large$|$}\!
  \sum_{\lambda\in P_k} \chii_\lambda(h_\alpha)\,\chii_\lambda(h)
  \mbox{\Large$|$}^2 \,,  \Labl2g
where $J$ is the weight factor in the Weyl integration formula
(see \Erf3a and \Erf a3) and $\caln$ is a normalization constant which 
is determined by the requirement that $\int_T\rmd h\,f(h)\eq1$. For
$\mathfrak{sl}(2)$, we have $T\eq[0,2]$ and $J(z)\eq\sin^2(\pi z)$.
The functions \Erf2g are then given by
  \be  f_{k,x}(z) = \caln_{k,x}\, \llb
  \sum_{\lambda=0}^k \sin((\lambda{+}1)\pi x)\,\sin(\pi z) {\lrb}^2  \Labl2u
for $z\iN[0,1)$, where $k\iN\zetplus$ and $x\eq(\mu{+}1)/(k{+}2)$ with
$\mu\iN\{0,1,...\,,k\}$.
The examples plotted in figure 1 are for conjugacy classes
$x\eq1/6$ and $1/2$ and for levels $k\eq4,\,10$ and $28$.

Closer inspection of the $\mathfrak{sl}(2)$ data also shows that
the sharpness of the localization scales with $k{+}\gv$. More
specifically, for any given conjugacy class $x$ and any
fraction $p$ of $\kappa\eq(k{+}2)^{-1}$, the integrated density 
$I_{x,p}\,{:=}\int_{x-\kappa/p}^{x+\kappa/p}\rmd z\,f(z)$ depends only 
very weakly on the level. In fact, we have collected extensive numerical 
evidence that even after taking this rescaling into account,
the localization improves when the level gets larger, i.e., that
$I_{x,p}(k)$ is monotonically increasing with $k$.
(The improvement is not spectacular, though. For instance,
$I_{x=\kappa,1}$ rises from $.9829$ at $k\eq3$ to $.9889$ at $k\eq20$, 
and $I_{x=\kappa,3}$ rises from $.6063$ to $.6074$.)

Note in particular that all brane world volumes are concentrated on {\em 
regular\/} conjugacy classes, and that already at small level the overlap 
with the exceptional conjugacy classes (i.e., $x\eq0$ and $x\eq1$ for
$\g\eq\mathfrak{sl}(2)$) is negligible. Indeed, as is clearly exhibited
by the last mentioned data, even the level-{\em dependent\/} allowed 
conjugacy classes that, at fixed level, are closest to an exceptional class 
(i.e., $x\eq\kappa$ for $\g\eq\mathfrak{sl}(2)$) are {\em not\/} driven
into the exceptional one in the infinite level limit. (Thus, in this 
respect, our findings do not agree with the prediction of the 
semi-classical analysis in \cite{alsc2} and in \cite{Gawe}. 
The origin of this discrepancy appears to be the absence of the shift 
$k\,{\mapsto}\,k{+}\gv$ in the classical setup. This shift occurs also 
naturally in other quantities like e.g.\ in character formul\ae\ and 
conformal weights.) This result will be confirmed by the investigation 
of the algebra of boundary operators, to which we now turn our attention.

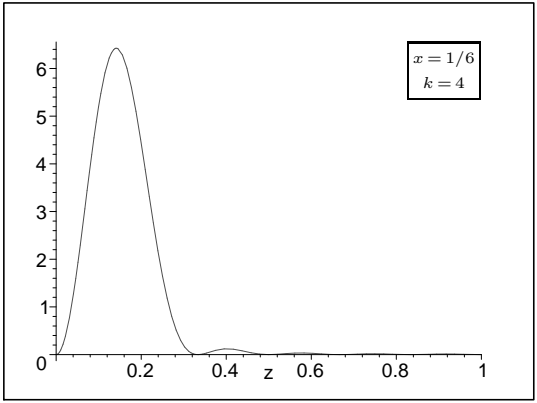
\begin{figure}[htbp]
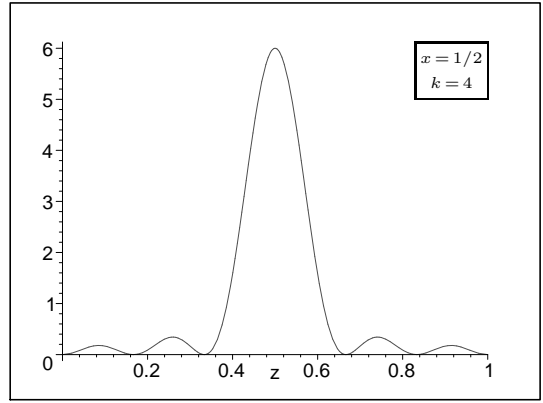
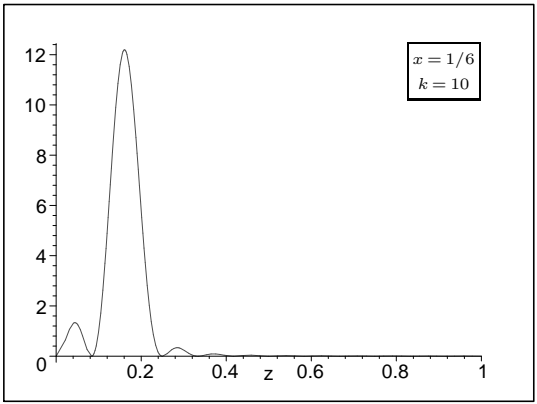
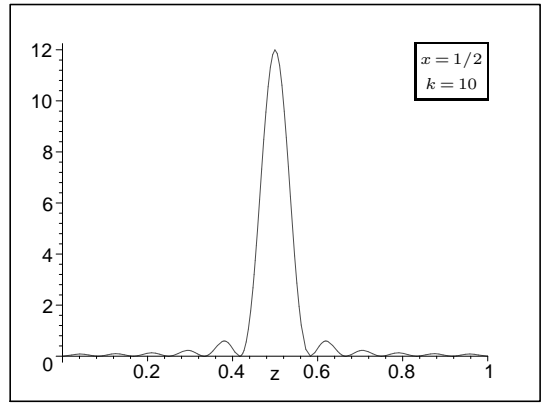
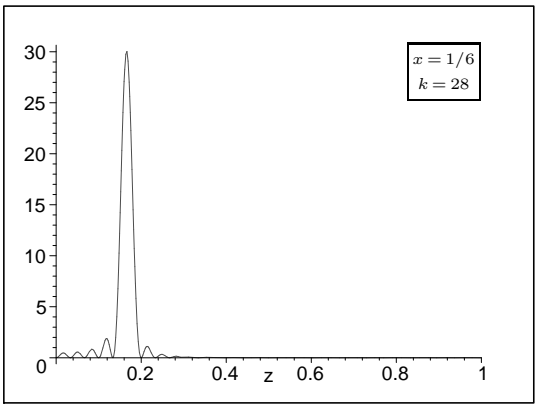
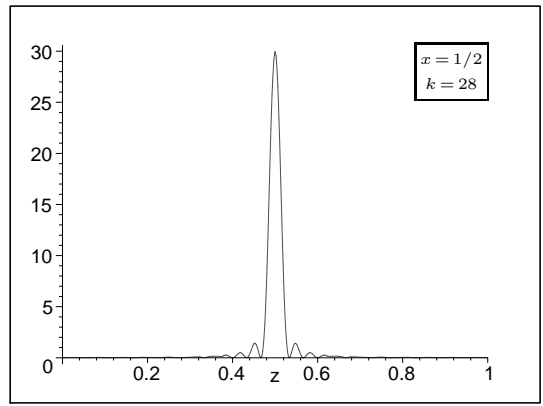
 \centering\mbox{
  \begin{tabular}{l}
\epsfig{file=fn6_4.psf,angle=-90,width=5.3cm} \hspace{8.5em}
\epsfig{file=fn2_4.psf,angle=-90,width=5.3cm}
\\[-2.3em]
\epsfig{file=fn6_10.psf,angle=-90,width=5.3cm} \hspace{8.5em}
\epsfig{file=fn2_10.psf,angle=-90,width=5.3cm}
\\[-2.3em]
\epsfig{file=fn6_28.psf,angle=-90,width=5.3cm} \hspace{8.5em}
\epsfig{file=fn2_28.psf,angle=-90,width=5.3cm}
  \end{tabular} \bPo(266,121)
\put(0,366){\infobox{1/6}{4}}  \put(259,366){\infobox{1/2}{4}}
\put(0,183){\infobox{1/6}{10}} \put(259,183){\infobox{1/2}{10}}
\put(0,0)  {\infobox{1/6}{28}} \put(259,0)  {\infobox{1/2}{28}} \eP
  }
\caption[] {The function \Erf2u for branes centered at the conjugacy 
classes $x\eq1/6$ and $x\eq1/2$ for levels $4,10,28$.}
\end{figure}

\sect{The algebra of boundary fields}

In this section we focus our attention on the operator product algebra of 
(WZW-primary) boundary fields. As a first basic ingredient, we need to 
determine by which quantum numbers such a field is characterized. Boundary 
points are precisely those points of a surface that have a unique pre-image
on its Schottky cover. Accordingly, on the level of chiral \cft, boundary 
operators are characterized by a single primary label $\lambda$.\,%
 \futnote{For WZW models, we are actually also interested in the
horizontal descendants, which have the same conformal weight as
the primary field. Then $\lambda$ should be regarded as a pair 
consisting of both the highest weight and the relevant actual $\gb$-weight.
Correspondingly, in the operator product \Erf tC below one must then 
in addition include the appropriate Clebsch\hy Gordan coefficients of $\gb$.}
 When all bulk symmetries are preserved, this label takes its values in the 
set of chiral bulk labels. (In the presence of symmetry breaking boundary
conditions, the analysis has to be refined, see \cite{fuSc11,fuSc12}.)
Moreover, a boundary operator typically changes the boundary condition;
therefore it carries two additional labels $\alpha,\beta$ which
indicate the two conformally invariant
boundary conditions at the two segments adjacent to the boundary insertion.

Boundary operators are therefore often written as
$\pso\lambda\beta\alpha{\!}(x)$. But, in fact, this is still not sufficient,
in general. The reason is that field-state correspondence requires to 
associate to {\em every\,} state that contributes to the partition function
  \be  {\rm A}_{\alpha\,\beta}(t) = \sum_\mu {\rm A}_{\alpha\,\beta}^\mu\,
  \chii_\mu(\Frac{\ii t}2) \ee
for an annulus with boundary conditions $\alpha$ and $\beta$ a separate
boundary field. For symmetry preserving boundary conditions, the
annulus coefficients ${\rm A}_{\alpha\,\beta}^\mu$ are known \cite{card9}\,%
 \futnote{Compare also \cite{Gawe} for arguments in a lagrangian setting.}
to coincide with fusion rule coefficients: 
  \be  {\rm A}_{\alpha\,\beta}^\mu = \N\beta\mu\alpha \,.  \Labl AN
The fusion rules $\N\beta\mu\alpha$ are not, in general, zero or one; as a
consequence one must introduce another degeneracy label $A$, taking values 
in $\{1,2,...\,,{\rm N}_{\alpha\P\beta\mu}\}$ \cite{fuSc11,Gawe,bppz2}.
The complete labelling of boundary operators therefore looks like 
  \be  \pso\lambda\beta\alpha A(x) \,.  \ee
We remark in passing that in more complicated situations,
like e.g.\ symmetry breaking boundary conditions \cite{fuSc11}
or non-trivial modular invariants \cite{fuSc5}, the degeneracy spaces
relevant for the boundary operators still admit a representation theoretic 
interpretation that involves appropriate (sub-)bundles of chiral blocks.

The boundary fields satisfy an operator product expansion which
schematically reads
  \be  \pso\lambda\alpha\beta A(x)\,\pso\mu\beta\gamma B(y)
  \,\sim\, \sum_\nu \sumn L\lambda\mu\nu \suma C \gamma\alpha\nu
  \tC{\lambda\mu}\nu\alpha\beta\gamma LABC \,
  [\pso \nu\alpha\gamma C(y) + ...\,] \,,  \Labl tC
where $L\iN\{1,2,...\,,\N\lambda\mu\nu\}$ labels a basis of the space
of chiral couplings from $\lambda$ and $\mu$ to $\nu$. 
It turns out that for every (rational) 
\cft, the structure constants $\tC{\lambda\mu}\nu\alpha\beta\gamma LABC$
appearing here are nothing but suitable entries of fusing matrices $\FF$:
  \be  \tC{\lambda\mu}\nu\alpha\beta\gamma LABC
  = \llb \F{\sxs L\nu C}{\sxs A{\beta\P}B}{\alpha\P}{\lambda\;}{\;\mu_{}}
  \gamma {\lrb}^* \,. \labl{tCtet}
Indeed, recall that the fusing matrices describe the transition between the 
$s$- and the $t$-channel of four-point blocks; pictorially, in our 
conventions, this relation looks like
\\[-2.9em]
  \be  \pTchannel0{66} \kappa\lambda\mu\nu\rho KL = \;
  \sum_\sigma \sumn M\kappa\lambda\sigma \sumn N\sigma\mu{\nu^+} \
  \F{\sxs K\rho L}{\sxs M\sigma N}\kappa\lambda\mu\nu   \ \cdot
  \pSchannel{11}{50} \kappa\lambda\mu\nu\sigma MN  \labl F
{}\\[2.1em]

To establish the identity \erf{tCtet}, we observe that the operator product 
coefficients $\tC{\lambda\mu}\nu\alpha\beta\gamma LABC$ furnish a solution 
of the sewing constraint \cite{lewe3,prss3} that arises from the two different 
factorizations of a correlation function of four boundary fields.
Including all degeneracy labels, this sewing relation reads
  \be  \bearl  \dsty \suma E \alpha\gamma\rho
  \tC{\kappa\lambda}\rho\alpha\beta\gamma KABE \,
  \tC{\mu\nu}{\rho\P}\gamma\delta\alpha LCD\EP \,
  \tCv\rho{\rho\P}\alpha\gamma\alpha E\EP
  \\{}\\[-1.4em] \hsp5
  = \dsty \sum_\sigma \sumn M \lambda\mu\sigma \sumn N \kappa\sigma{\nu\P}
  \suma F \beta\delta\sigma \tC{\lambda\mu}\sigma\beta\gamma\delta MBCF \,
  \tC{\kappa\sigma}{\nu\P}\alpha\beta\delta NAF\DP \,
  \tCv{\nu\P}\nu\alpha\delta\alpha\DP D
  \cdot \F{\sxs M\sigma N}{\sxs K\rho L}\kappa\lambda\mu\nu \,,  \eear \Labl4b
where ${\sf F}\Barray{\lambda\,\mu}{\kappa\,\nu}$ is the fusing matrix
which relates the two different factorizations.
For WZW models, the fusing matrices coincide with the $6j$-symbols of 
the corresponding quantum group with deformation parameter
a $k{+}\gv$th root of unity. The constraint \Erf4b can be solved 
explicitly in full generality, without reference to the particular \cft\ 
under investigation.  First note that the structure constants 
$\tC{\lambda\mu}\nu\alpha\beta\gamma LABC$ depend on six chiral and four 
degeneracy labels, so that their label structure is precisely the same as the 
one of the fusing matrices. The key observation is then to realize the 
similarity between the constraint \Erf4b and the pentagon identity
  \be \bearl \dsty \sumn \EP \gamma\rho\alpha
        \F{\sxs\KP{\kappa\P}A}{\sxs B{\gamma\P}\EP}{\beta\P}{\lambda\;}
                                                               {\rho\P}{\alpha\;}
  \cdot \F{\sxs\LP{\rho\P}\EP}{\sxs C{\delta\P} D} {\gamma\P}{\mu\;}{\;\nu}\alpha
  \\{}\\[-1.6em] \hsp5
  = \dsty \sum_\sigma \sumn M \lambda\mu\sigma \sumn N \sigma\nu{\kappa\P}
                      \sumn F \delta\sigma\beta
        \F{\sxs M\sigma F}    {\sxs B{\gamma\P}C} {\beta\P}{\lambda\;}{\,\mu}\delta
  \cdot \F{\sxs N{\kappa\P}A} {\sxs F{\delta\P}D} {\beta\P}{\sigma\;}{\,\nu}\alpha
  \cdot \F{\sxs\LP{\rho\P}\KP}{\sxs M\sigma N} \lambda{\mu\,}\nu\kappa
  \eear  \labl5
for the fusing matrices. 

The identification \erf{tCtet} was already deduced before in \cite{bppz2} 
from the structural similarity between the factorization 
constraint \Erf4b and the pentagon identity. (For the case of Virasoro minimal 
models, this identification had been observed even earlier \cite{runk2} by
using special properties of those models.) 
Indeed, it is not too difficult to show that under the identification
\erf{tCtet} the constraint \Erf4b becomes -- after exploiting the tetrahedral 
($S_4$-) symmetry \cite{fgsv} of the fusing matrices in order to replace\,%
 \futnote{Note that the tetrahedral transformations that do not preserve
the orientation of the tetrahedron involve complex conjugation of $\FF$.
Also, by the tetrahedral symmetry, the structure constants involving the
vacuum label $\Omega$ are just combinations of quantum dimensions, and
these precisely cancel against the quantum dimensions coming from the
other tetrahedral transformations that have to be performed.
More details will be given elsewhere.}
 some of the fusing matrix entries by different ones --
nothing but the (complex conjugated) pentagon identity \erf5.
In this context, we would like to stress that the fusing matrices are 
entirely defined in terms of {\em chiral\/} \cft.  

A more direct way to understand the result \erf{tCtet} is by interpreting the
boundary fields $\pso\lambda\alpha\beta A$ as (ordinary) chiral
vertex operators, which pictorially amounts to the prescription
\\[-4.1em]
  \be  \pso\lambda\alpha\beta A \,\ \hat= \hsp{.9} \bP(85,80)(0,11)
    \putlin0010{60} \putlin{30}001{30} \putvec{14}0{-1}0
    \putvec{51}0{-1}0 \putvec{30}{22}0{-1}
    \putsc{27.8}{33.5}{\lambda} \putsc{3.1}{-5.9}{\alpha}
    \putsc{52.3}{-7.3}{\beta}
    \putss{27.1}{-6.1}{A}
  \eP \ee
{}\\[1.3em] The operator product \Erf tC then describes the transition\\[-3.3em]
  \be \bP(103,50)(0,11)
    \putlin0010{90} \putlin{30}001{30} \putlin{60}001{30}
    \putvec{14}0{-1}0 \putvec{46}0{-1}0 \putvec{81}0{-1}0
    \putvec{30}{22}0{-1} \putvec{60}{22}0{-1}
    \putsc{27.8}{33.5}{\lambda} \putsc{57.6}{33.9}{\mu}
    \putsc{3.1}{-5.9}{\alpha} \putsc{45.3}{-7.3}{\beta}
    \putsc{82.3}{-5.9}{\gamma}
    \putss{27.1}{-5.9}{A} \putss{57.1}{-5.9}{B}
  \eP \longrightarrow \bP(85,80)(0,40)
    \putlin{20}{20}10{60} \putlin{50}{50}{-1}1{22}
    \putlin{50}{50}11{22} \putlin{50}{50}0{-1}{30}
    \putvec{30}{70}1{-1} \putvec{70}{70}{-1}{-1} \putvec{50}{36}0{-1}
    \putvec{34}{20}{-1}0 \putvec{71}{20}{-1}0
    \putsc{21.3}{75}{\lambda} \putsc{74.5}{75}{\mu}
    \putsc{51.4}{37.6}{\nu}
    \putss{47.8}{53.3}{L} \putss{47.5}{13.9}{C}
    \putsc{23.4}{14.2}{\alpha} \putsc{72.3}{14.2}{\gamma}
  \eP \ee
{}\\[1.7em]
from which one can read off the desired identification \erf{tCtet} 
between boundary structure constants and fusing matrices.

Thus we conclude that the boundary structure constants are indeed nothing 
but suitable entries of fusing matrices.
It should be noted, however, that the fusing matrices are not
completely determined by the pentagon identity and their tetrahedral
symmetry. Rather, there is a gauge freedom related to the possibility
of performing a change of basis in the spaces of chiral three-point couplings.
 In the present setting, the gauge invariance corresponds to the freedom in 
choosing a basis in the space of all boundary fields $\pso\lambda\alpha\beta A$ 
with fixed $\lambda$ and $\alpha,\beta$. Once the gauge freedom has been fixed
at the level of chiral three-point couplings, it is natural to
make the same gauge choice also for the boundary fields.

\medskip
We now show that, upon appropriately taking the limit of infinite 
level, the algebra of those boundary operators that do not change the boundary 
condition approaches the algebra $\calf(G/T)$ of functions on 
the homogeneous space $G/T$, where $T$ is a maximal torus of the 
Lie group $G$. This space $G/T$ is of interest to us because every
regular conjugacy class of $G$ is, as a differentiable manifold, isomorphic
to $G/T$. Our result therefore perfectly matches the fact that the fusing 
matrices of \wzwm s can be expressed with the help of $k{+}\gv$th roots 
of unity, i.e.\ again the level $k$ gets shifted by the dual Coxeter number 
$\gv$. Correspondingly, the weights are shifted by the Weyl vector so that, 
again, we are naturally led to regular conjugacy classes.

We start our argument by regarding the algebra $\calf(G/T)$ as a left
$G$-module only, rather than as an algebra. The module $\calf(G/T)$ is fully
reducible and can be decomposed as follows. According to \Erf1i, the space 
$\calf(G)$ of functions on $G$ is a $G$-bimodule under left and right 
translation. Since the right action of $T$ on $G$ is free, we can 
then identify $\calf(G/T)$ with the subspace of $T$-invariant functions on $G$,
  \be  \calf(G/T) = \calf(G)^T \,\cong\, \bigoplus_{\lambda\in P}
  \calhb_\lambda^{} \otimeS \llb \calhb_{\lambda\PP}{\lrb}^T \, . \ee
Furthermore, invariance under the maximal torus $T$
just picks the weight space for weight zero. Thus we find
  \be  \calf(G/T) \,\cong\, \bigoplus_{\lambda\in P}\m_0^{(\lambda)}\,
  \calhb_\lambda  \labl{ds}
as an isomorphism of $\gb$-modules, where $\m_0^{(\lambda)}$ is the 
multiplicity of the weight $\mu\eq0$ in the irreducible module 
$\calhb_\lambda$ with highest weight $\lambda$. Thus,
in particular, only modules belonging to the trivial conjugacy class
of $\gb$-modules appear in the decomposition \Erf ds.
Recall that the boundary operators are organised in terms of modules
$\calh_\lambda$ of the affine \lie\ $\g$. Our aim is to show that the algebra 
of boundary fields that correspond to states of lowest grade in the modules 
$\calh_\lambda$, i.e.\ which are either primary fields or horizontal 
descendants,\,%
 \futnote{In general \cfts, there is no underlying `horizontal' structure,
unlike in WZW models. It is not clear whether, in general, it is the set of
states of lowest conformal weight, or e.g.\ the quotient (called `special 
subspace') of $\calh_\lambda$ that was introduced in \cite{nahm8},
that is relevant in this argument.
But already for \wzwm s this truncation is not sufficiently well understood.}
 carries a left $G$-module structure
that in the limit of infinite level coincides with the decomposition \Erf ds.

When restricting to this finite subspace $F_k$ of boundary operators,
via field-state correspondence the annulus amplitudes tell us that
$F_k$ carries the structure of
a $G$-module, and as a $G$-module it is isomorphic to the direct sum
  \be  F_k \,\cong\, \bigoplus_{\mu,\nu\in P_k} {\rm A}_{\mu\,\nu}^\lambda\,
  \calhb_\lambda  \Labl Fk
of irreducible $G$-modules. Thus to be able to perform a 
more quantitative analysis of the algebra of boundary 
operators, we need to control the values of the annulus coefficients
${\rm A}_{\mu\,\nu}^\lambda$. Since according to the identity \Erf AN, as 
long as all bulk symmetries are preserved, these numbers just coincide with
the fusion rules coefficients, $A_{\mu\nu}^\lambda\eq {\rm N}_
{\mu\P\nu\lambda}$, we are interested in concrete expressions for the 
fusion rules.  It turns out that they can be expressed through suitable 
weight multiplicities in the following convenient form:
  \be  {\rm N}_{\mu\nu^+\lambda} = \Frac1{|W|} 
  \sum_{w_1,w_2\in W}\! \eps(w_1)\,\eps(w_2) \sum_{\beta\in\Lv} 
  \m^{(\lambda)}_{-w_1(\mu+\rho)+w_2(\nu+\rho)-\beta(k+\gv)} \,.  \labl{rel}
Here $W$ is the Weyl group of $\gb$ and $\eps$ its sign function, and
the sum over $\beta$ extends over the coroot lattice $\LV$ of $\gb$. The
relation \erf{rel} can be derived by combining the Kac\hy Walton formula 
for WZW fusion coefficients with Weyl's character formula
and Weyl's integration formula. For details, see appendix \ref{AA}. 

We are interested in the behavior of the numbers \erf{rel} in the limit of
large $k$. As for boundary conditions, this limit must be taken with care. 
As we have seen, they can be labelled either by conjugacy classes of group
elements $h_\alpha$ or by the corresponding $\gb$-weights $\alpha$. But the 
relation \erf{rcw} between these two types of data involves explicitly
the level $k$, so that we have to decide which of the two is to be kept
fixed in the limit. In the present
context, we keep the conjugacy classes fixed. Accordingly we consider 
two sequences, denoted by $\mu_k^{}$ and $\nu_k^{}$, of weights such that
  \be  y_\mu := \Frac{\mu_k^{} +\rho}{k+\gv} \quad{\rm and}\quad
  y_\nu := \Frac{\nu_k^{} +\rho}{k+\gv}  \ee
do not depend on $k$ (and $\exp(2\pi\ii y_\mu)$ and $\exp(2\pi\ii y_\nu)$
are regular elements of the maximal torus $T$ of $G$).

In terms of these quantities, in formula \erf{rel} the multiplicity of 
the weight 
  \be  \mu_k^{} := (k{+}\gv)\, (-w_1(y_\mu)+w_2(y_\nu) - \beta) \ee
appears, with fixed $y_\mu$ and $y_\nu$ (and fixed $w_1,w_2$ and $\beta$). 
At large $k$ this weight becomes larger than any
non-zero weight of the module $\calhb_\lambda$, except when the relation
  \be  -w_1(y_\mu)+w_2(y_\nu) - \beta = 0 \Labl;;
is satisfied. As a consequence, at large level,
the action of element $w_2$ of the Weyl group $W$ of $\gb$
on the regular element $y_\nu$ must coincide with 
the action of the element $(w_1,\beta)$ of the corresponding affine Weyl 
group $\hat W$ on $y_\mu$. This, however, is only possible when 
$y_\mu\eq y_\nu$, and then it follows that $\beta\eq 0$ as well as 
$w_1\eq w_2$. Thus the requirement \Erf;; has $|W|$ many solutions. 
We thus obtain in the limit of infinite level
  \be  \lim_{k\to\infty} {\rm N}_{\mu_k^+\nu_k^{}\lambda} =
  \delta_{y_\mu,y_\nu}\, \m^{(\lambda)}_0 \,.  \ee

In view of the relation between fusion rules and annulus coefficients we 
thus learn that, in the limit of large level, only those pairs of boundary
conditions contribute which correspond to identical conjugacy classes, or 
in other words, only those open strings survive which start and end at the
same conjugacy class. (For every finite value of $k$, however, such open
strings are still present.) Moreover, in this limit the non-vanishing
annulus partition functions become
  \be  \lim_{k\to\infty} {\rm A}_{y_\mu\,y_\mu}(t) = \sum_{\lambda\in P}
  \m^{(\lambda)}_0\, \chii_\lambda({\ii t}/2) \,,  \Labl1l
so that \Erf Fk simplifies to
  \be  \lim_{k\to\infty} F_k \,\cong\, \bigoplus_{\lambda\in P} 
  \m^{(\lambda)}_0\, \calhb_\lambda \,.  \Labl Fi
This space is indeed nothing but $\calf(G/T)$ as appearing in \erf{ds}. 
Our result indicates in particular that the algebra of boundary
operators that do not change the boundary condition is related to the space
of {\em functions\/} on the brane world volume. This should be regarded as
empirical evidence for a statement that is not obvious in itself, since in 
general non-trivial vector bundles over the brane can appear as Chan\hy Paton 
bundles, so that boundary operators might as well be related
to sections of non-trivial bundles rather than to functions.

In summary, in the limit of infinite level those boundary operators that
belong to states in the \findim\ subspace
$\calhb_\lambda\,{\subseteq}\,\calh_\lambda$ of lowest conformal weight
furnish a $G$-module that is isomorphic to the algebra $\calf(G/T)$
of functions on a regular conjugacy class, seen as a $G$-module. 

So far we have considered the spaces of our interest only as $G$-modules. 
But we would like to equip both $\calf(G)$ and the space of 
boundary operators also with an algebra structure. The operator product 
algebra of boundary operators, whose structure constants
are, as we have seen, fusing matrices, obeys certain associativity
properties. These properties are not immediately related to ordinary
associativity, because the definition of the operator product
involves a limiting procedure.

Several proposals have been made recently for the relation between 
the operator product algebra of boundary operators
and the algebra $\calf(G/T)$. An approach based on deformation 
quantization was proposed in \cite{gapl}. The definition of the product
then involves fixing the insertion points of the two boundary fields at 
prescribed positions in parameter space.
As the theory in question is {\em not\/} topological, one is thus forced
to introduce arbitrary and non-intrinsic data -- in contrast to the 
situation with topological theories studied in \cite{cafe}. Another proposal 
\cite{alrs3} starts from a restriction of the operator product \alg\
to fields that correspond to the states of lowest conformal weight in the
affine irreducible modules. This destroys associativity.
 The prescription in \cite{alrs3} also allows only for open strings that 
have both end points on one and the same brane. This is difficult to reconcile 
with the fact that (compare formula \Erf1l above) open strings connecting 
different branes can only be ignored in the limit of infinite level.

\sect{Symmetry breaking boundary conditions}

We now turn to boundary conditions of \wzwm s that break part of the bulk 
symmetries. One important class of consistent \bc s can be constructed by 
prescribing
an automorphism $\omega$ of the chiral algebra that connects left
movers and right movers in the presence of a boundary. In this case the 
boundary condition is said to have {\em automorphism type\/} $\omega$. We 
point out, however, that also boundary conditions are known for which no 
such automorphism exists. A WZW example is provided by $\mathfrak{so}(5)$ 
at level 1; in this example, there is a conformal embedding
with a subalgebra isomorphic to $\mathfrak{sl}(2)$ at level 10, and one
can classify boundary conditions (see \cite{bppz2}) that preserve only
the $\mathfrak{sl}(2)$ symmetries. However, no general theory for
such boundary conditions without automorphism type has been developped so far, 
and we will not consider them in the present paper.

Every \bc\ preserves some subalgebra $\calap$ of the full chiral algebra
$\cala$; because of conformal invariance, $\calap$ contains the Virasoro
sub\alg\ of $\cala$. For \bc s that do possess an automorphism type $\omega$, 
the preserved subalgebra $\calap\,{\subseteq}\,\cala$ can be characterized as 
an orbifold subalgebra, namely as the algebra $\calap\eq\cala^{{<}\omega{>}}$
consisting of elements that are fixed under $\omega$. A theory treating such 
boundary conditions for arbitrary \cfts\  has been developped in 
\cite{fuSc11,fuSc12}. In the case of interest to us, the relevant
automorphisms of the chiral algebra are induced by automorphisms $\omega$
of the horizontal subalgebra $\gb$ of the untwisted affine \lie\ $\g$ that
preserve the compact real form of $\gb$. Via the construction \erf{aff}
of affine \lie s as centrally extended loop algebras, every such
automorphism $\omega$ extends uniquely to an automorphism of $\g$.
By a slight abuse of notation we denote this automorphism by $\omega$, too.

We briefly summarize some of the results that we will derive in this 
section. In the same way that symmetry preserving boundary conditions are
localized at regular conjugacy classes, the boundary conditions of automorphism
type $\omega$ are localized at the submanifolds
  \be  \calc_G^\omega(h) := \{ g h \omega(g)^{-1} \,|\, g\iN G \} \,,  \ee
to which we refer as twined conjugacy classes, with $h\iN G$ of the form 
\erf{rcw}.\,%
 \futnote{Note that when $\omega$ is an involution, then the twined conjugacy 
class $\calc_G^\omega(e)$ of the identity element $e$ of $G$ is the symmetric
space $G/G^\omega$. Because of the shift $k\,{\mapsto}\,k{+}\gv$ the
element $e\iN G$ is not  of the form \erf{rcw}, hence this space
does not correspond to any boundary condition of the WZW model.}
 In the case of $\mathfrak{sl}(2)$ or, more generally, whenever $\omega$ is an 
inner automorphism of $\g$, the twined conjugacy classes are just tilted
versions of ordinary conjugacy classes. 
 More precisely, they can be obtained from ordinary conjugacy classes by right 
translation, $\calc_G^{\Ad_s}(h)\eq\calc_G^{}(hs)\,s^{-1}$. 

In the case of outer automorphisms, the dimension of twined conjugacy classes 
differs from the dimension of ordinary ones. While ordinary regular
conjugacy classes are isomorphic to the homogeneous space $G/T$, twined
conjugacy classes for outer automorphisms turn out to be isomorphic
to $G/T_0^\omega$, where $T_0^\omega$ is a subtorus of the maximal torus $T$.
For instance, for $\g\eq\mathfrak{sl}(3)$ the dimension of regular conjugacy
classes is $\dim(G/T)\eq8{-}2\eq6$, while for outer automorphisms twined 
conjugacy classes have dimension $\dim (G/T_0^\omega)\eq8{-}1\eq7$. The 
increase in the dimension actually generalizes a well-known
effect in free \cfts, where all automorphisms are outer, to the non-abelian 
case. Namely, in a flat $d$-dimensional background the relevant automorphism,
which is an element of ${\rm O}(d)$, determines the dimension of the brane and 
a constant field strength on it. In particular, non-trivial automorphisms 
can change the dimension of the brane.
\medskip

The boundary states $\calb_\alpha^\omega$ for symmetry breaking boundary 
conditions of automorphism type $\omega$ are built from {\em twisted 
boundary blocks\/} $B^\omega_\lambda$ \cite{fuSc12}. For the latter, 
the Ward identities \erf1 get generalized to
  \be  B^\omega_\lambda \circ \left( J^a_{n}\oT\bfe 
  + \bfe\oT \omega(J^a_{-n}) \right) = 0 \,.  \Labl t1
To proceed, we need some further information on automorphisms of $\gb$ 
that preserve the compact real form. Such automorphisms are
in one-to-one correspondence to automorphisms of the connected
and simply connected compact real Lie group $G$ whose \lie\ is the
compact real form of $\gb$. For each such automorphism
$\omega$ there is a maximal torus $T$ of $G$ that is invariant under $\omega$.
The complexification $\mathfrak t$ of the \lie\ of $T$ is a \csa\ of $\gb$. 
The torus $T$ is not necessarily pointwise fixed under $\omega$. The subgroup
  \be  T^\omega := \{ t\iN T \,|\, \omega(t)\eq t \}  \ee
of $T$ that {\em is\/} left pointwise fixed under $\omega$ can have
several connected components \cite{wend,wend2}. The connected component of 
the identity will be denoted by $T^\omega_0$.

The automorphism $\omega$ of $\gb$ can be written as the composition of 
an inner automorphism, given by the adjoint action $\Ad_s$ of some element 
$s\iN T$, with a diagram automorphism $\omego$: 
  \be \omega(g) = \omego(sgs^{-1}) \,;  \labl{descr}
without loss of generality, $s$ can be chosen to be invariant under 
$\omega$, $\omega(s)\eq s$.  Let us recall the definition of a 
{\em diagram automorphism\/}. Any symmetry $\dotomego$ of 
the \dyd\ of $\gb$ induces a permutation of the root generators $E^i_\pm$ 
that correspond to the simple roots of $\gb$ with respect to the Cartan 
subalgebra $\mathfrak t$, according to
  \be  E^i_\pm \;\mapsto\; \omego(E^i_\pm) := E^{\dotomego i}_\pm \,.  \ee
This extends uniquely to an automorphism $\omego$ of $\gb$ that preserves 
the compact real form and is called a diagram automorphism of $\gb$. 
When $\omega$ is an inner automorphism
then the diagram automorphism in the decomposition \erf{descr} is the 
identity; in general, $\omego$ accounts for the outer part of $\omega$. Also 
note that, for inner automorphisms, $T^\omega$ is the full maximal torus $T$.

As $\omega$ leaves a \csa\ $\mathfrak t$ invariant, there is an associated 
dual map $\omegas$ on the weight space $\mathfrak t^\star$ of $\gb$. Applying 
the condition \erf{t1} for the zero modes, i.e.\
$n\eq0$, one sees that non-zero twisted boundary 
blocks only exist for {\em symmetric weights\/}, i.e.\ weights $\lambda$
satisfying $\omegas(\lambda)\eq\lambda$. Note that relation \erf{descr} 
implies that $\omegas(\lambda)\eq\omega_\circ^\star(\lambda)$, so that in the
case of inner automorphisms all integrable highest weights $\lambda$ contribute.

Next, we explain what the coefficients
in the expansion of the symmetry breaking boundary states \wrtt twisted
boundary blocks are, i.e., what the correct generalization of the numbers
$S_{\lambda,\alpha}/S_{\Omega,\alpha}$ appearing in formula \Erf,1
is. We have seen in \Erf25 that for $\omega\eq\id$, these coefficients are 
given by the characters $\chii_\lambda$ of $G$, evaluated at specific elements 
\erf{rcw} of the maximal torus $T$. For general $\omega$, the analogous 
numbers have been determined in \cite{bifs}. For the present purposes it 
is most convenient to express them as so-called {\em twining characters\/} 
\cite{fusS3,furs}, evaluated at specific elements $h_\alpha$ of $T$.

Let us explain what a twining character is. To any automorphism $\omega$
of $\gb$ we can associate twisted intertwiners $\Theta_\omega$, that is,
linear maps 
  \be \Theta_\omega:\quad \calhb_\lambda \to \calhb_{\omegas\lambda} \ee
between $\gb$-modules that obey the twisted intertwining property
  \be \Theta_\omega \circ R_\lambda(x) = R_{\omegas\lambda}(\omega(x)) \circ
  \Theta_\omega \labl{twtw}
for all $x\iN\gb$. By Schur's lemma, the twisted intertwiners are unique
up to a scalar. For symmetric weights, $\omegas(\lambda)\eq\lambda$, 
the twisted intertwiner $\Theta_\omega$ is an endomorphism. In this case we 
fix the normalization of $\Theta_\omega$ by requiring that $\Theta_\omega$ 
acts as the identity on the highest weight vector. 
For symmetric weights, the twining character $\chii^\omega_\lambda$
is now defined as the generalized character-valued index
  \be  \chii^\omega_\lambda(h) := \tr_{\calhb_\lambda} \Theta_\omega 
  R_\lambda(h) \,.  \labl{twin}
Character formul\ae\ for twining characters of arbitrary (generalized)
\kma s have been established in \cite{fusS3,furs}. 

Finally we describe at which group elements $g\iN G$ the twining character
must be evaluated in order to yield the coefficients of the boundary state.
The integral $\gb$-weights form a lattice $\Lw$ consisting of all
elements of $\mathfrak t^\star$ of the form
$\lambda\eq\sum_{i=1}^{{\rm rank}\,\bar{\mathfrak g}} \lambda^i\Lambda_{(i)}$
that obey $\lambda^i\iN\zet$ for all $i$. Both the Weyl group $W$ and the 
automorphism $\omegas$ act on this lattice. We will also need 
a lattice $\LVomegas$ that contains the lattice $\Lwomega$
of integral symmetric $\gb$-weights, i.e., of integral weights
satisfying $\omegas(\lambda)\eq\lambda$ or, equivalently,
$\lambda^{\dotomego i}\eq\lambda^i$ for all $i\eq1,2,...\,,{\rm rank}\,\gb$. 
The lattice $\LVomegas$ consists of symmetric $\gb$-weights as well, but
we weaken the integrality requirement by imposing only that 
$N_i\lambda^i\iN\zet$ for 
all $i$. Here $N_i$ denotes the length of the corresponding orbit of the \dyd\ 
symmetry $\dotomego$. For brevity we call this lattice $\LVomegas$ the 
lattice of {\em fractional symmetric weights\/}. By construction, the lattice 
$\LVomegas$ is already determined uniquely by the outer \auto\ class of 
$\omega$. In particular, when $\omega$ is inner, then both $\LVomegas$ and the 
symmetric weight lattice $\Lwomega$ just coincide with the ordinary weight 
lattice $\Lw$.

The lattice $\Lw$ of integral weights of $\gb$ has as a sublattice
the lattice $\LV$ of integral linear combinations 
  \be  \beta=\sum_{i=1}^{{\rm rank}\,\bar{\mathfrak g}} \beta_i\,
  \alpha^{(i)\Vee}  \ee
of simple coroots $\alpha^{(i)\Vee}$. In analogy to what we did before for 
weights, we also introduce another lattice $\Lwomegas$, the lattice of 
{\em fractional symmetric coroots\/}, by requiring that $\omegas(\beta)
\eq\beta$ and $N_i\beta_i\iN\zet$ for all $i$. We have the inclusions
$\LVomega\,{\subseteq}\,\Lwomegas$ and $\Lwomega\,{\subseteq}\,\LVomegas$.

On both the lattice $\LVomegas$ of fractional symmetric weights and 
the lattice $\Lwomegas$ of fractional symmetric coroots, we have
an action of a natural subgroup $W^\omega$ of the Weyl group $W$, namely
of the commutant
  \be  W^\omega := \{ w\iN W \,|\, w\omegas\eq\omegas w \} \,.  \Labl Wo
The group $W^\omega$ depends only on the diagram part $\omego$ of $\omega$;
in particular, for inner automorphisms, $\omegas$ is the identity and hence 
$W^\omega\eq W$. For outer \auto s, $W^\omega$ can be described explicitly 
\cite{furs} as follows. For the outer automorphisms of $A_{2n}$, $W^\omega$ is 
isomorphic to the Weyl group of $C_n$; for $A_{2n+1}$ to the Weyl group of 
$B_{n+1}$; for $D_n$ to the one of $C_{n-1}$; and for $E_6$ to the Weyl group of
$F_4$. Finally, for the diagram automorphism of order three of $D_4$ one obtains
the Weyl group of $G_2$. (This whole structure allows for a generalization
to arbitrary \kma s, and the commutant of the Weyl group can be shown to be
the Weyl group of some other \kma, the so-called orbit \lie\ \cite{fusS3}.)
The group $W^\omega$ also acts on the fixed subgroup $T^\omega$ of the maximal
torus $T$. One can show that the twining characters \erf{twin} are invariant
under the action of $W^\omega$, which generalizes the invariance of ordinary
characters under the full Weyl group $W$.

To characterize the symmetry breaking boundary conditions, we now choose some
fractional symmetric weight $\alpha\iN\LVomegas$.
It is not hard to see that the group element
  \be  h_\alpha := \exp(2\pi\ii y_\alpha) \,,  \labl{4.26}
where $y_\alpha$ is the corresponding dual element in the \csa, i.e.\
$y_\alpha\,{:=}\,\Frac{\alpha+\rho}{k+\gv}$, depends on $\alpha$ only
modulo fractional symmetric coroots. Moreover, the subgroup $W^\omega$ of 
the Weyl group $W$ acts freely on the set of all $h_\alpha$;
there are as many different orbits as there are symmetric integrable weights.
Accordingly, we should actually regard the label $\alpha$ of a \bc\ of
\auto\ type $\omega$ as an element
  \be  \alpha \in \LVomegas /
  \llb W^\omega{\ltimes}(k{+}\gv)\Lwomegas\lrb \,.  \ee
A \bc\ is then uniquely characterized by an element of this finite set.
Letting $\alpha$ run over this set, we obtain all conformally invariant
boundary conditions of automorphism type $\omega$.

Let us list a few other properties of the group element $h_\alpha$. It is an 
element of the fixed subgroup $T^\omega$ of the maximal torus, or more 
precisely, of the connected component $T^\omega_0$ of
the identity of $T^\omega$. Moreover, it is a regular element of $G$.

Furthermore, it should be mentioned that in the special case of outer automorphisms 
of $\gb\eq A_{2n}$, there is an additional subtlety in the description of
the twined conjugacy classes. It arises from
the fact \cite{bifs} that in this case the extension of the diagram 
automorphism of $\gb$ to the affine \lie\ $\g$ does not exactly give the diagram
automorphism of $\g$. The additional inner automorphism of $\g$ is taken
into account by the adjoint action of an appropriate element $s_\circ$ of the
maximal torus. Namely, denote by $x_\circ$ the dual of the weight
$\frac14(\Lambda_{(n)}{+}\Lambda_{(n+1)})$, i.e.\ the \csa\ element such that
$(x_\circ,x)\eq\frac14(\Lambda_{(n)}{+}\Lambda_{(n+1)})(x)$ for all $x$ in the
\csa\ of $\gb$. Then, for outer automorphisms of $A_{2n}$, formula \erf{4.26} 
must be generalized to
  \be  h_\alpha := \exp(2\pi\ii y_\alpha)\, \exp(2\pi\ii x_\circ)  \,.  \ee

We are now finally in a position to write down the boundary states explicitly;
we have
  \be  \calb^\omega_\alpha = \sum_{\lambda\in P_k^\omega} 
  \chii^\omega_\lambda(h_\alpha)\, B^\omega_\lambda  \Labl,2
with $P_k^\omega$ the set of symmetric weights in $P_k^{}$.
For trivial automorphism type, $\omega\eq\id$, we recover formula \erf{,1}.

Fortunately, all the group theoretical tools that we used in the previous
sections have generalizations to the case of twining characters (for details
see appendix \ref{AB}). Therefore,
once we have expressed the boundary states in the form \erf{,2}, we are also
able to generalize the statements of sections 2 and 3 to the case of
symmetry breaking boundary conditions. For instance, recall that
ordinary characters are class functions, 
  \be  \chii_\lambda(g h g^{-1}) = \chii_\lambda(h) \,,   \ee
i.e.\ they are constant on conjugacy classes $\calc_G$ \Erf cg.
Combining the cyclic invariance of the trace and the twisted intertwining
property \erf{twtw} of the maps $\Theta_\omega$, one learns that twining 
characters are {\em twined class functions\/} in the sense that 
  \be  \chii^\omega_\lambda(g h\, \omega(g)^{-1}) = \chii^\omega_\lambda(h)
  \,.  \labl{twcl}
As a consequence, the {\em twined conjugacy classes\/}
  \be   \calc_G^\omega(h) := \{ g h \omega(g)^{-1} \,|\, g\iN G \}  \ee
and the {\em twined adjoint action\/}
  \be  \Ad^\omega_g:\quad h \;\mapsto\; g\, h\, \omega(g)^{-1}  \Labl ao
(i.e.\ the twined version of the adjoint action $\Ad^{}_g$ of $g\iN G$)
will play exactly the roles for symmetry breaking boundary conditions 
that ordinary conjugacy classes $\calc_G(h)$ and ordinary adjoint action 
$\Ad^{}_g$ play in the case 
of symmetry preserving boundary conditions. We refrain from presenting 
details of the calculations; for some hints and for the necessary group 
theoretical tools, such as a twined version of Weyl's
integration formula, we refer to appendix \ref{AB}.

We summarize a few properties of twined conjugacy classes (for details see
appendix \ref{AB}). Every group element $g\iN G$ can be mapped
by a suitable twined adjoint map to $T^\omega_0$. For regular elements 
$h\iN G$, the twined conjugacy class is isomorphic, as a manifold with
$G$-action, to the homogeneous space
  \be \calc_G^\omega(h) \cong G / T^\omega_0 \, . \ee

For outer automorphisms, the following intuition appears to be accurate. The
twined conjugacy classes are submanifolds of $G$ of higher dimension. To 
characterize them by the intersection\,%
 \futnote{One word of warning is, however, in order. The orbits of twined 
conjugation
intersect $T^\omega_0$ in several points, but, in contrast to the standard
group theoretical situation, the intersections are not necessarily related 
by the action of $W^\omega$. Rather, a certain extension $W(T^\omega_0)$ of 
$W^\omega$, to be described in appendix \ref{AB}, is needed \cite{wend,wend2}.}
 with elements of the maximal torus, it is therefore sufficient to restrict 
to the symmetric part $T^\omega$ of the maximal torus (and even to the 
connected component $T^\omega_0$ of it). In contrast, for an inner automorphism 
$\omega\eq\Ad_s$ with $s\iN G$, the twined conjugacy classes have the 
same shape as ordinary conjugacy classes; indeed, they are just obtained 
by right-translation of ordinary conjugacy classes:
  \be  \calc_G^{\Ad_s} (h) = \calc_G^{}(hs)\, s^{-1}  \, . \ee

The twined analogue of the formula \Erf;; requires only the symmetric part 
of the weight to vanish (because in the twined analogue of \erf{,A} 
only equality of the symmetric parts of the weights is enforced by the
integration). As a consequence, at fixed \auto\ type $\omega$ the large
level limit
\erf{Fi} of the boundary operators gets replaced by
  \be  \lim_{k\to\infty} F^\omega_k \,\cong\, \bigoplus_{\lambda\in P} 
  \m^{(\lambda)}_{0,\omega} \, \calhb_\lambda \,,  \ee
where $\m^{(\lambda)}_{0,\omega}$ stands for the sum of the
dimensions of all weight spaces of $\calhb_\lambda$ for weights whose
symmetric part vanishes. The limit $\lim_{k\to\infty} F^\omega_k$ again 
yields the algebra of functions on the brane world volume which in this 
case is isomorphic, as a manifold, to the homogeneous space $G/T^\omega_0$.

\sect{Non-simply connected group manifolds}

In this section we extend the results of the previous two sections to 
Lie group manifolds $G$ that are not simply connected. Before we 
present our results in more detail, we briefly outline them for the group
$G\eq\SO(3)$. As is well-known, $\SO(3)$ is obtained as the quotient of 
the simply connected group $\SU(2)$ by its center $\zet_2$. We will see that
to every symmetry preserving boundary condition for $\SO(3)$ we can again
associate a conjugacy class of $\SO(3)$. The latter are projections of 
orbits of conjugacy classes of the covering group $\SU(2)$ under the action 
of the center $\zet_2$. Thinking of the group manifold $\SU(2)$ as the
three-sphere $S^3$ with the north pole being the identity element $+\one$
and the south pole the non-trivial element $-\one$ of the center, the action
of the center is the antipodal map on $S^3$. The conjugacy classes that are 
related by the center are then those having the same `latitude' on $S^3$. 
Those conjugacy classes which describe
boundary conditions must obey the same integrality constraints 
as in the $\SU(2)$ theory. Explicitly, at level $k$ the two 
$\SU(2)$ conjugacy classes $(\lambda{+}\rho)/(k{+}\gv)$ and 
$(k{-}\lambda{+}\rho)/(k{+}\gv)$ give rise to a single boundary condition for
$\SO(3)$. An additional complication arises for the `equatorial' conjugacy 
class $\lambda\eq k/2$, which is invariant under the action of the center;
it gives rise to two distinct boundary conditions. Also note that all 
automorphisms of $\SO(3)$ are inner, and thus in one-to-one correspondence
with automorphisms of $\SU(2)$. Symmetry breaking boundary conditions of 
$\SO(3)$ therefore correspond to tilted $\SO(3)$ conjugacy classes.

This picture is reminiscent of the phenomena one encounters in orbifold 
theories, and indeed the \wzwt\ based on the group $\SO(3)$ can be understood
\cite{fegk,fegk1} as an orbifold of the $\SU(2)$ \wzwt. Branes of the orbifold
theory correspond to symmetric brane configurations in the covering space;
branes at fixed point sets give rise to several distinct boundary conditions,
known as `fractional branes' \cite{didg}. We point out, however, the following
additional feature that is revealed by our analysis. Namely, in case the 
orbifold group admits non-trivial two-cocycles, branes at fixed point sets 
do not necessarily split.  To what extent a splitting occurs is controlled 
by the cohomology class of the relevant two-cocycles.
\smallskip

Let us now describe our results more explicitly. For the time being, 
we restrict our attention to boundary conditions that preserve all 
bulk symmetries. The compact connected simple Lie group $G$
can be written as the quotient of a simply connected, compact and 
connected universal covering group $\tilde G$ by an appropriate subgroup
$\Gamma$ of the center of $\tilde G$. There is a natural projection
  \be  \pi:\quad \tilde G \to G  \Labl pi
whose kernel is the finite group
$\Gamma$. As a consequence, the \wzwt\ based on $G$ can be
seen as an `orbifold' of the theory based on $\tilde G$. (It should be
pointed out, however, that the term `orbifold' is used in this context
in a broader sense than is commonly done in the \rep\ theoretic formulation
of orbifolds in \cft, compare e.g.\ to \cite{dvvv}.)

It is known \cite{fegk,fegk1} that the \wzwt\ on a non-simply connected group 
manifold is described by a non-diagonal modular invariant that can
be constructed with the help of simple currents. The relevant simple
currents are in one-to-one correspondence with the elements of
the subgroup $\Gamma$ of the center of $\tilde G$. In the most general
situation, the non-diagonal modular invariant in question is obtained
by applying a so-called simple current automorphism to a chiral \cft\ 
that is itself constructed from the original diagonal theory by a 
simple current extension \cite{scya6}. 
For the sake of simplicity, in the sequel we will discuss only such
conformally invariant boundary conditions for which only one of
the two mechanisms, i.e., either a simple current automorphism or a
simple current extension, is present. For $\tilde G\eq\SU(2)$, 
{\em both\/} cases correspond to the non-simply connected quotient 
$\SO(3)\eq\SU(2)/\zet_2$; the former arises for levels of the form 
$k\eq2\bmod 4\zet$, where one deals with a modular invariant of $D_{{\rm 
odd}}$-type, while the latter appears for levels $k\eq0\bmod 4\zet$ 
and corresponds to a modular invariant of $D_{{\rm even}}$-type.

We first consider simple current extensions. We can then invoke 
the general result that boundary conditions preserving all bulk symmetries 
are labelled by the primary fields of the relevant \cft, which is now not the
\wzwt\ corresponding to $\tilde G$, but the \cft\ that is obtained from it by
the simple current extension. This extended theory can be 
described as follows \cite{fusS6}. Its primary fields correspond to certain
orbits of the action of $\Gamma$ on the primary fields of the unextended
theory. But only a certain subset of orbits is allowed, e.g.\ for
$G\eq\SO(3)$ only those that correspond to integer spin highest weights. 
We will see later, however, that the other orbits describe conformally 
invariant boundary conditions as well. Those boundary conditions do
not preserve all symmetries of the extended chiral algebra, but they still
preserve all symmetries of the chiral algebra for the $\tilde G$-theory.

We also must account for the fact that the action of $\Gamma$ on the set 
of orbits is not necessarily free.\,%
 \futnote{While the (left or right) action of $\Gamma$ on individual group 
elements is obviously free, the action on conjugacy classes can be non-free,
since $h$ and $\eps h$ with $\eps\iN\Gamma$ can belong to the same 
conjugacy class.}
 When it is not free, then there are several distinct primary fields associated 
to the same orbit. For determining the number of primaries coming from such an 
orbit, one must take into account the fact that the action of the simple 
current group is in general only projective; an algorithm for solving this
problem has been developped in \cite{fusS6}. We summarize these findings 
in the statement that the boundary conditions of the \wzwt\ based on
$G$ correspond to orbits of conjugacy classes of $\tilde G$ under the
action of $\Gamma$, with multiplicities when this action is not free.

Next, we study the case of automorphism modular invariants. For this
situation the boundary conditions that preserve all bulk symmetries 
have been found in \cite{prss3} for $\tilde G\eq\SU(2)$
and in \cite{fuSc5} for the general case. They are labelled by orbits of the
action of $\Gamma$ on primary fields, or, equivalently, on conjugacy classes.
Again, when this action is not free, then there are several
inequivalent boundary conditions associated to the same orbit. On disks
with boundary conditions that come from the same orbit, bulk fields in the 
untwisted sector possess identical one-point functions, but the one-point
functions of bulk fields in the twisted sector are different for
different \bc s of this type.  They differ in sign, and the absolute values 
are controlled by the matrices $S^{\rm J}$ that describe the modular 
S-transformation of one-point chiral blocks on the torus with insertion 
of the relevant simple currents $\rm J$ \cite{fuSc5}. 

To provide a geometrical interpretation of these results, we first
relate conjugacy classes of the group $G$ to conjugacy classes of its
covering group $\tilde G$.  The conjugacy class $\calc_{G}(h)$ of an element 
$h\iN G$ in the non-simply connected group $G$ can be written as the image 
under the map $\pi$ \Erf pi of several conjugacy classes
$\calc_{\tilde G}$ of the universal covering group $\tilde G$. We claim that
  \be  \pi^{-1}(\calc_G(h)) = \bigcup_{\eps\in\Gamma} \,
  \calc_{\tilde G}(\eps \tilde h) \,,  \labl{nscon}
where $\tilde h\iN\tilde G$ is any a lift of $h$, i.e.\ $\pi(\tilde h)\eq h$. 
To see that the set on the right hand side of \erf{nscon} is contained in 
the set on the left hand side, we note that its elements are of the form
$\tilde g\, \eps \tilde h\, \tilde g^{-1}$ for some $\tilde g\iN \tilde G$
and some $\eps\iN\Gamma$. Further, we have
  \be  \pi(\tilde g\, \eps \tilde h\, \tilde g^{-1})
  = \pi(\tilde g)\, \pi(\tilde h)\, \pi(\tilde g^{-1}) = g h g^{-1} \ee
where $g$ is the projection $\pi(\tilde g)$; since $g h g^{-1}$ lies
in $\calc_G(h)$, indeed $\tilde g\, \eps \tilde h\, \tilde g^{-1}$ 
is contained in the left hand side of \erf{nscon}. 
Conversely, assume that $h'\iN G$ is conjugate to $h\iN G$, which means 
that $ghg^{-1}\eq h'$ for some $g\iN G$. There exists a $\tilde g\iN
\tilde G$ such that $\pi(\tilde g)\eq g$, and every element of 
$\pi^{-1}(ghg^{-1})$ is of the form
$(\eps_1\tilde g)\,(\eps_2\tilde h)\,(\eps_3\tilde g^{-1})$ for suitable
elements $\eps_1,\eps_2,\eps_3\iN\Gamma$. Using that the $\eps_i$ are
central in $\tilde G$, this means that $\pi^{-1}(h')$ lies in the set on
the right hand side of \erf{nscon}.

Let us now consider those conjugacy classes which are left invariant by some
subgroup $\Gamma'$ of $\Gamma$. (For example, the group manifold 
$\tilde G\eq\SU(2)$ is a three-sphere $S^3$, and the regular conjugacy classes
are isomorphic to spheres $S^2$ of fixed latitude; thus there is a single
conjugacy class that is fixed by the action of the center $\zet_2$ of 
$\tilde G$, namely the equatorial conjugacy class. At level $k$, 
it corresponds to the weight $\mu\eq k/2$ that is a fixed point with respect 
to fusion with the non-trivial simple current of the $\mathfrak{sl}(2)$
\wzwt.) The finite subgroup $\Gamma'$ acts freely on such an
invariant conjugacy class $\calco$. Therefore the space $\calf(\calco)$ 
of functions on $\calco$ can be decomposed into eigenspaces under the action 
of $\Gamma'$. In the simplest case, the subspaces just consist of odd and 
even functions, respectively. In general, the decomposition reads
  \be \calf(\calco) = \bigoplus_{\psi\in{\Gamma'}^*} \calf_{\!\psi}(\calco)
  \,,  \labl{ns1}
where the eigenvalues $\psi$ are given by characters of $\Gamma'$.

It follows that the boundary conditions for non-simply connected groups $G$ 
can be described by conjugacy classes of $G$ itself, with the important
subtlety that those conjugacy classes which are invariant under the action 
of the group $\Gamma$ give rise to several distinct boundary conditions.
Our analysis reproduces, in particular, the following familiar features 
of D-branes on orbifold spaces. Brane configurations on the original 
space $\tilde G$ that are symmetric under the action of $\Gamma$
give rise to boundary conditions in the quotient $G$.
Individual branes that are invariant under a subgroup
$\Gamma'$ of the orbifold group $\Gamma$ yield several boundary
conditions which differ in the contribution from the twisted sector;
the coefficients in their boundary states are reduced by a common factor,
which is precisely the effect of fractional branes \cite{didg}.

We can also describe the analogue of the decomposition \erf{ns1} of functions
on invariant branes for boundary operators. Again, we discuss simple current
extensions and automorphisms separately. In the case of automorphisms, it
was shown in \cite{fuSc5} that the annulus multiplicities are given by
the rank of the sub-vector bundle of chiral blocks with definite 
parity under the simple current automorphism. In the case of simple
current extensions, the annulus multiplicities are, according to \cite{card9},
fusion rules of the $\tilde G$-theory. Moreover, general results
\cite{fusS6} on the fusion rules of a simple current extension show that
the fusion rules of the extended theory -- that is, in our case, of
the $G$-theory -- are given by sub-bundles of definite parity as well.
Just like for simply connected groups, our analysis therefore confirms the 
general idea that the algebra of boundary operators should be a quantization 
of the algebra of functions on the brane world volume.

We also would like to point out one important subtlety in the analysis of
invariant orbits. The exact analysis \cite{fuSc11} reveals that not all
invariant orbits necessarily split off and give rise to several boundary
conditions. Rather, it can happen that the action of the stabilizer of the 
orbifold group in the underlying orbifold construction is only projective,
and in this case even an invariant conjugacy class can give rise to only
a single boundary condition. An example is given by 
$\tilde G\eq{\rm Spin}(8)/\zet_2{\times\zet_2}$; at level 2, there is a single
conjugacy class that is fixed under $\Gamma$, and yet, due to the appearance 
of a genuine {\em untwisted stabilizer\/} \cite{fusS6}, it gives rise
to a single conformally invariant boundary condition. For more details,
we refer to \cite{fuSc12}.

We proceed to briefly discussing some aspects of symmetry breaking boundary 
conditions for \wzwts\ on non-simply connected group manifolds. 
We first discuss which automorphisms can be used. While every automorphism of 
$\gb$ that preserves the compact real form gives rise to an automorphism of 
the universal covering group $\tilde G$, such automorphisms do not
necessarily descend to the quotient group $G$. Rather, every automorphism
of $\tilde G$ restricts to an automorphism of the center $\calz(\tilde G)$ of 
$\tilde G$; for an inner automorphism this restriction is the 
identity. The automorphisms of $\tilde G$ that descend to automorphisms of 
$G\eq\tilde G/\Gamma$ are precisely those that map $\Gamma$ to itself.
Notice that the group of {\em inner} automorphisms of $\tilde G$ and $G$
coincide; in both cases this group is equal to the adjoint group
$\tilde G/\calz(\tilde G) \,{\cong}\, G/\calz(G)$. 
The symmetry breaking boundary conditions for non-simply connected group 
manifolds that come from automorphisms are therefore related to twined
conjugacy classes of $G$ in much the same way as in the simply connected
case, with the same subtleties arising for twined conjugacy classes that are
left invariant by some element of the center.

We finally remark that in the case of extensions, such as those for $A_1$ 
at level $k\eq0\bmod 4$, another type of symmetry breaking boundary 
condition exists for the $G$-theory, namely boundary conditions which only 
preserve the symmetries of the unextended theory, i.e.\ of the 
$\tilde G$-theory. These come from \auto s of the extended chiral \alg\ 
that act as the identity on the unextended one. It has been demonstrated 
\cite{fuSc11} that such boundary conditions are labelled by
$\tilde G$-primaries as well. 
As already mentioned, they correspond to those $\Gamma$-orbits of conjugacy 
classes of $\tilde G$ that are projected out in the $\tilde G$ theory.
For $G\eq\SO(3)$, for instance, they are obtained by projection from conjugacy
classes of $SU(2)$ that are related to half-integer spin highest weights. 
We can therefore describe also this type of boundary conditions by orbits of 
$\tilde G$-conjugacy classes which by \erf{nscon} project, in turn,
to $G$-conjugacy classes.

\vskip3em\noindent{\small {\bf Acknowledgement} \\
We would like to thank A.\ Lerda, K.\ Gaw\c edzki, O.\ Grandjean, and
B.\ Pedrini for stimulating discussions.}
 
     \newpage

\appendix

\sect{Fusion rules}\label{AA}

In this appendix we derive the relation \erf{rel} between fusion rule
coefficients and weight multiplicities. We start with the observation
that a character can, on one hand, be written in terms of weight multiplicities
  \be \chii_\lambda (h) = \sum_\mu m^{(\lambda)}_\mu\, \eE^{\mu}(h) \,,  \ee
and on the other hand can be expressed in terms of Weyl's character formula as
  \be \chii_\lambda (h) = X^{-1}(h) \sum_{w\in W} \eps(w)\,
  \eE^{w(\lambda+\rho)}(h) \,. \labl{wchar}
Here the sum is over the Weyl group $W$ of $\gb$, $\eps$ is the sign function
on $W$, and 
  \be X(h) := \eE^{\rho}(h) \prod_{\alpha>0} (1-\eE^{-\alpha}(h))  \Labl3a
is the well-known expression for the denominator. 
(Up to an exponential $\eE^{\lambda+\rho}$, $X^{-1}$ is just the character 
of the corresponding Verma module of highest weight $\lambda$.)

Next we recall the Kac\hy Walton formula \cite{KAc3,walt+} for WZW fusion 
rules. It expresses the fusion coefficients ${\rm N}_{\mu\nu\lambda}$ as an 
alternating sum over a certain subset $\Wp$ of the 
affine Weyl group $\hat W$. $\Wp$ consists by definition of those elements
of $\hat W$ that map the fundamental Weyl alcove to some alcove in the
fundamental Weyl chamber. The set $\Wp$ furnishes a 
distinguished set of representatives for the coset $\hat W/ W$, but
$\Wp$ is not a group. The representatives can be characterized by the fact
that they have minimal length. The Kac\hy Walton rule yields
  \be  {\rm N}_{\mu\nu^+\lambda} = \sum_{\wP\in\WP} \eps(\wp)\, 
  \call_{\wP(\mu+\rho)-\rho,\nu^+\!,\lambda} \,,  \labl{kawa}
where $\call_{\wP(\mu),\nu^+\!,\lambda}$ is the dimension of the space of 
singlets in the tensor product $\calhb_{\wP(\mu)}\otimeS \calhb_{\nu^+}\otimeS
\calhb_\lambda$ of the three $\gb$-modules $\calhb_{\wP(\mu)}$, 
$\calhb_{\nu^+}$ and $\calhb_\lambda$. This dimension, in turn, can be 
expressed in terms of an integral over the corresponding characters as
  \be \bearll  \call_{\wP(\mu+\rho)-\rho,\nu^+\!,\lambda} \!\!
  &= \dsty \int_G\! \rmd g\; \chii_{\wP(\mu+\rho)-\rho}(g)\,
  \chii_{\nu^+}(g)\, \chii_\lambda(g)
  \\{}\\[-.8em]
  &= \dsty \Frac1{|W|} \int_T\! \rmd h\, J(h)\, \chii_{\wP(\mu+\rho)-\rho}
  (h)\, \chii_{\nu^+}(h)\, \chii_\lambda(h) 
  \,,  \end{array}\labl{a1}
where in the second line we have used Weyl's integral formula to reduce the
integral to an integral over a maximal torus $T$ of $G$.

The next step is to insert the formula \Erf a1 into the Kac-Walton rule
\erf{kawa} and to
recombine the summations over $W$ and $\Wp$. At the same time, we use the 
Weyl character formula to rewrite the characters $\chii_{\wP(\mu+\rho)-\rho}$
and $\chii_{\nu^+}$, while the third character is expressed in terms of weight 
multiplicities. We then arrive at
  \be \hsp{-.8}  \bearll  {\rm N}_{\mu\nu^+\lambda} \!\!
  &= \dsty \Frac1{|W|} \int_T\! \rmd h\, \sum_{\wP\in\WP} \eps(\wp)\, J(h)\,
  \chii_{\wP(\mu+\rho)-\rho}(h)\, \chii_{\nu^+}(h)\, \chii_\lambda(h)
  \\{}\\[-.8em]
  &= \dsty \Frac1{|W|} \sum_{w_1,w_2\in W} \eps(w_1)\eps(w_2)
  \sum_{\beta\in\Lv} \int_T\! \rmd h\, \eE^{w_1(\mu+\rho)-w_2(\nu+\rho)
  +(k+\gv)\beta}(h)\, \chii_\lambda(h)
  \\{}\\[-.8em]
  &= \dsty \Frac1{|W|} \sum_{w_1,w_2\in W} \eps(w_1)\eps(w_2) 
  \sum_\sigma \sum_{\beta\in\Lv}
  \int_T\! \rmd h\, \eE^{w_1(\mu+\rho)-w_2(\nu+\rho)+(k+\gv)\beta}(h)\,
  \m^{(\lambda)}_\sigma\, \eE^\sigma(h)
  \\{}\\[-.8em]
  &= \dsty \Frac1{|W|} \sum_{w_1,w_2\in W}  \eps(w_1)\eps(w_2)
  \sum_{\beta\in\Lv} \m^{(\lambda)}_{-w_1(\mu+\rho)+w_2(\nu+\rho)-(k+\gv)\beta}
  \,, \end{array} \Labl,A
so that we have finally arrived at the relation \erf{rel}.
Here in the second line we have also used the following two simple 
relations. First, the characters of two conjugate modules are related as
  \be \chii_{\lambda\PP}(h) = \chii_\lambda (h^{-1}) \,.  \labl{a2}
Second, the Jacobian factor $J$ in Weyl's integration formula can be
expressed in terms of $X$ as 
  \be  J(h) = X(h)\, X(h^{-1}) \,.  \labl{a3}
Together they allow us to cancel the two Weyl denominators against the
volume factor $J$. The $\sigma$-summation in the third line of \Erf,A is
over the weight system of $\calhb_\lambda$, and in the last line the 
integral over the maximal torus $T$ was evaluated explicitly. 

\sect{Twined conjugation}\label{AB}

To investigate the properties of the twined conjugation \erf{ao}, it turns 
out to be helpful to relate it to the theory of non-connected Lie groups. 
The non-connected Lie groups for which the connected component of the identity
is isomorphic to a given real, compact, connected and simply connected Lie 
group $G$ can be related to subgroups of the group of automorphisms
of the Dynkin diagram of the \lie\ $\gb$ whose compact real form is the
\lie\ of the group $G$. (This should not be confused with the 
relation between non-{\em simply\/} connected groups and automorphisms of the
{\em extended\/} \dyd.) Namely, for every subgroup $\check\Gamma$ of 
diagram automorphisms of $\gb$, one can construct a Lie group $\check\G$ with 
the group of connected components given by $\pi_0(\check G)\eq\check \Gamma$ 
as the semi-direct product of the Lie group $G$ and the finite group 
$\check\Gamma$. Conversely, if $g$ is any element of a Lie group $\check G$
that is not in the connected component of the identity, then the adjoint action 
of $g$ on the \lie\ $\gb$ is given by an {\em outer\/} automorphism $\omega_g$ 
and therefore corresponds to a symmetry of the \dyd\ of $\gb$.

The non-trivial connected components of $\check G$ are, as differentiable 
manifolds with metric, isomorphic to $G$. We fix a connected component
$G_{\dot\omega}$ that corresponds to the element $\dot\omega$ of the group of 
\dyd\ symmetries of $\gb$. The adjoint action of any element $g\iN G$ of the 
connected component of the identity maps $G_{\dot\omega}$ to itself. Taking any
arbitrary element $g_{\dot\omega}\iN G_{\dot\omega}$, we can write every element
in $G_{\dot\omega}$ as $h g_{\dot\omega}$ with $h\iN G$, 
and we have $\omega(g)\eq g_{\dot\omega}^{} g g_{\dot\omega}^{-1}$.
For the adjoint action of $g\iN G$ we then find
  \be  \Ad_g(h g_{\dot\omega}) = g\, h\, \omega(g)^{-1}\, g_{\dot\omega}
  = \Ad_g^\omega(h)\, g_{\dot\omega}  \ee
with $\omega\,{\equiv}\,\omega_{g_{\dot\omega}}$.
We see that, after choosing an origin $g_{\dot\omega}$ for $G_{\dot\omega}$,
ordinary conjugation by $g\iN G$ acts on $h$ like twined conjugation.
Changing the origin $g_{\dot\omega}$ changes the relevant automorphism by an 
inner automorphism.

Now denote by 
  \be  N^\omega(T^\omega_0) := \{ g\iN G \,|\, gt\omega(g)^{-1} \iN T^\omega_0
  \mbox{ for all } t\in T^\omega_0 \} \ee
the {\em twined normalizer\/} of the connected component $T^\omega_0$ in the 
fixed subgroup of the maximal torus $T$. The quotient
  \be  W(T^\omega_0) := N(T^\omega_0) / T^\omega_0 \ee
is called the Weyl group of $T^\omega_0$. It can be shown \cite{wend,wend2}
that $W(T^\omega_0)$ is the product of the subgroup $W^\omega$ of the Weyl 
group that was defined in \Erf Wo and a finite abelian group $\Gamma(G,\omega)$.
Moreover, the mapping degree of the mapping
  \be \bearll
  q_\omega: \ & G/T^\omega_0 \times T^\omega_0 \;\to\; G 
  \\{}\\[-.8em]&
  (g T^\omega_0, t) \;\mapsto\;  g t \omega(g)^{-1} 
  \end{array}\ee
is \cite{wend,wend2} $\,\deg q_\omega\eq|W(T^\omega_0)|$. In particular, the 
mapping degree is positive, so $q_\omega$ is surjective. This, in turn, implies 
that any group element of $G$ can be mapped by a suitable twined conjugation 
\erf{ao} into $T^\omega_0$, which generalizes the well-known conjugation 
theorems for the maximal torus.

The determinant of $q_\omega$ can be computed. One finds at the
point $(1,h)$ with $h\iN T^\omega_0$
  \be \det q_\omega = |\Gamma(G,\omega)|\, \mbox{\Large$|$}\!
  \prod_{\check\alpha>0} \llb 1- \eE^{2\pi\ii \check\alpha}(h) \lrb
  \mbox{\Large$|$}^2 =: |\Gamma(G,\omega)| \cdot J^\omega(h) \,,   \ee
where the product is over a set of weights that are constructed from 
$\omegas$-orbits of positive $\gb$-roots and which can be shown \cite{fusS3} 
to be isomorphic to the set of positive roots of the so-called orbit \lie\
that is associated to $\gb$ and $\omega$. (Recall that $W^\omega$
is isomorphic to the Weyl group of the orbit \lie.)
Application of Fubini's theorem then yields the twined generalization 
  \be  \int_G\! \rmd g\, f(g) = 
  \Frac1{|W^\omega|} \int_{T^\omega_0}\!\! \rmd h\, J^\omega(h)\;
  \llb \int_{G/T^\omega_0} \!\!\rmd(gT^\omega_0)\, f(g t \omega(g)^{-1}) \lrb  \ee
of Weyl's integration formula.
Here $\rmd g$, $\rmd h$ and $\rmd(g T^\omega_0)$ are the Haar measures
on the Lie groups $G$ and $T^\omega_0$ and on the homogeneous space
$G/T^\omega_0$, respectively. Obviously, the integration formula is 
particularly useful for twined class functions $\chii^\omega$ (see 
\erf{twcl}), for which it reduces to 
  \be  \int_G\! \rmd g\, \chii^\omega(g) =
  \Frac1{|W^\omega|} \int_{T^\omega_0}\! \rmd h\, J^\omega(h)\, \chii^\omega(h) 
  \,.  \ee

 \vskip3em
\small
 \newcommand\wb{\,\linebreak[0]} \def\wB {$\,$\wb}
 \newcommand\Bi[1]    {\bibitem{#1}}
 \newcommand\J[5]   {{\sl #5}, {#1} {#2} ({#3}) {#4}}
 \newcommand\PhD[2]   {{\sl #2}, Ph.D.\ thesis (#1)}
 \newcommand\Prep[2]  {{\sl #2}, preprint {#1}}
 \newcommand\BOOK[4]  {{\em #1\/} ({#2}, {#3} {#4})}
 \newcommand\Dipl[2]  {{\sl #2}, Diploma thesis (#1)}
 \def\jf    {J.\ Fuchs}
 \newcommand\inBO[7]  {in:\ {\em #1}, {#2}\ ({#3}, {#4} {#5}), p.\ {#6}}
 \newcommand\inBOx[7] {in:\ {\em #1}, {#2}\ ({#3}, {#4} {#5})}
 \newcommand\gxxI[2] {\inBO{GROUP21 Physical Applications and Mathematical
              Aspects of Geometry, Groups, and \A s{\rm, Vol.\,2}}
              {H.-D.\ Doebner, W.\ Scherer, and C.\ Schulte, eds.}
              \WS\Si{1997} {{#1}}{{#2}}}
 \def\anop  {Ann.\wb Phys.}
 \def\atmp  {Adv.\wb Theor.\wb Math.\wb Phys.}
 \def\comp  {Com\-mun.\wb Math.\wb Phys.}
 \def\duke  {Duke\wB Math.\wb J.}
 \def\duki  {Duke\wB Math.\wb J.\ (Int.\wb Math.\wb Res.\wb Notes)}
 \def\foph  {Fortschr.\wb Phys.}
 \def\ijmb  {Int.\wb J.\wb Mod.\wb Phys.\ B}
 \def\ijmp  {Int.\wb J.\wb Mod.\wb Phys.\ A}
 \def\imrn  {Int.\wb Math.\wb Res.\wb Notices}
 \def\jams  {J.\wb Amer.\wb Math.\wb Soc.}
 \def\jhep  {J.\wb High\wB Energy\wB Phys.}
 \def\jomp  {J.\wb Math.\wb Phys.}
 \def\joal  {J.\wB Al\-ge\-bra}
 \def\jopa  {J.\wb Phys.\ A}
 \def\jram  {J.\wB rei\-ne\wB an\-gew.\wb Math.}
 \def\nupb  {Nucl.\wb Phys.\ B}
 \def\phlb  {Phys.\wb Lett.\ B}
 \def\phrd  {Phys.\wb Rev.\ D}
 \def\phrl  {Phys.\wb Rev.\wb Lett.}
 \def\thmp  {Theor.\wb Math.\wb Phys.}
 \newcommand\geap[2] {\inBO{Physics and Geometry} {J.E.\ Andersen, H.\
            Pedersen, and A.\ Swann, eds.} \MD\NY{1997} {{#1}}{{#2}} }
 \def\AMS    {{American Mathematical Society}}
 \def\AP     {{Academic Press}}
 \def\CUP    {{Cambridge University Press}}
 \def\MD     {{Marcel Dekker}}
 \def\NH     {{North Holland Publishing Company}}
 \def\SV     {{Sprin\-ger Ver\-lag}}
 \def\WS     {{World Scientific}}
 \def\Ad     {{Amsterdam}}
 \def\Be     {{Berlin}}
 \def\Ca     {{Cambridge}}
 \def\NY     {{New York}}
 \def\PR     {{Providence}}
 \def\Si     {{Singapore}}

\end{document}